\documentclass[12pt, a4paper]{article}
\pdfoutput=1
\usepackage{jcappub} % for details on the use of the package, please
\usepackage{datetime}
\usepackage{amsmath}
\usepackage{amsfonts}
\usepackage{amssymb}
\usepackage{latexsym}
\usepackage{graphicx}
\usepackage{color}
\usepackage{mathrsfs}
\usepackage{slashed}
\usepackage{hyperref}
\usepackage{enumerate}
\usepackage{multirow}
\usepackage[table]{xcolor}
\usepackage{colortbl}
\usepackage{url}
\usepackage{enumitem}
\usepackage{comment}
\usepackage{bm}
\usepackage{float}

\numberwithin{equation}{section}

\hypersetup{colorlinks, citecolor=bluscuro, linkcolor=magenta, urlcolor=bluscuro}
\definecolor{rossos}{rgb}{0.7,0,0.3}
\definecolor{violachiaro}{rgb}{1,0.6,1}
\definecolor{rossochiaro}{rgb}{1,0.6,0.6}
\definecolor{verdechiaro}{rgb}{0.6,1,0.6}
\definecolor{giallochiaro}{rgb}{1,1,0.6}
\definecolor{bluscuro}{rgb}{0.15, 0.2, 0.9}
\definecolor{verdes}{rgb}{0.1, 0.5, 0.1}
\definecolor{gold}{rgb}{1,0.84,0}
\definecolor{forestgreen}{rgb}{0.13,0.55,0.13}
\makeatother   % Cancel the effect of \makeatletter

\newcommand{\fraction}[2]{\frac{\displaystyle #1}{\displaystyle
#2}}

 \def\be   {\begin{equation}}   \def\ee   {\end{equation}}
 \def\ba   {\begin{array}}      \def\ea   {\end{array}}
 \def\bea  {\begin{eqnarray}}   \def\eea  {\end{eqnarray}}
 \def\bean {\begin{eqnarray*}}  \def\eean {\end{eqnarray*}}

\def\myurl#1#2{\href{http://#1}{#2}}

\begin{document}
%\hfill\textit{\currenttime\qquad \dmyyyydate\today}

\title{Antiproton constraints on the GeV gamma-ray excess: a comprehensive analysis}

\author[a]{Marco Cirelli}
\affiliation[a]{Institut de Physique Th\'eorique, CNRS URA2306 \& CEA-Saclay,\\ 91191 Gif-sur-Yvette (France)}

\author[b,c]{Daniele Gaggero}
\affiliation[b]{SISSA, via Bonomea 265, 34136 Trieste (Italy)}
\affiliation[c]{INFN, Sezione di Trieste, via Valerio 2, 34127 Trieste (Italy)}

\author[a]{Ga\"elle Giesen}

\author[a]{Marco Taoso}

\author[b]{Alfredo Urbano}

\emailAdd{marco.cirelli@cea.fr}
\emailAdd{daniele.gaggero@sissa.it}
\emailAdd{gaelle.giesen@cea.fr}
\emailAdd{marco.taoso@cea.fr}
\emailAdd{alfredo.urbano@sissa.it}


\proceeding{\small Saclay-T14/091\\ SISSA  41/2014/FISI} 

\date{\today}

%\begin{abstract}
\abstract{A GeV gamma-ray excess has possibly been individuated in {\sc Fermi-LAT} data from the Galactic Center and interpreted in terms of Dark Matter (DM) annihilations, either in hadronic (essentially $b\bar{b}$) or leptonic channels. In order to test this tantalizing interpretation, we address two issues: (i) we improve the computation of secondary emission from DM (Inverse Compton and Bremsstrahlung) with respect to previous works, confirming it to be very relevant for determining the DM spectrum in the leptonic channels, so that any conclusion on the DM nature of the signal critically depends on this contribution; (ii) we consider the constraints from antiprotons on the DM hadronic channel, finding that the uncertainties on the propagation model, and in particular on the halo height, play a major role. Moreover, we discuss the role of solar modulation, taking into account possible charge dependent effects whose importance is estimated exploiting detailed numerical tools. 
The limits that we obtain severely constrain the DM interpretation of the excess in the hadronic channel, for standard assumptions on the Galactic propagation parameters and solar modulation. However, they considerably relax if more conservative choices are adopted.}
%The constraints that we obtain are relevant for the DM interpretation of the excess in the hadronic channel, but less stringent with respect to other similar %analyses, as a consequence of the more conservative approach.}

%\end{abstract}

\maketitle
\flushbottom

\section{Introduction}

Due to the very high density of Dark Matter (DM) particles that is predicted in the inner part of the Galaxy, the Galactic Center (GC) region is expected to be one of the brightest sources of radiation coming from DM annihilation or decay. For this reason, the very accurate gamma-ray maps and spectra provided by the {\sc Fermi-LAT} collaboration have been investigated in detail by the DM community, in order to find some hint of an excess with respect to the expected emission from astrophysical processes. The complexity of the region and the abundance of sources make however the task of separating the emission by DM annihilation/decay from backgrounds extremely challenging. In spite of that, several authors have reported since 2009 the detection of a gamma-ray signal from the inner few degrees around the GC~\cite{hooperon_history,Abazajian:2014fta}. Its spectrum and morphology are claimed to be compatible with those expected from annihilating DM particles: to fix the ideas, we recall the results of the most recent analysis~\cite{Daylan:2014rsa}, which confirms the presence of this excess at an incredibly high level of significance (if taken at face value) and finds this signal to be best fit by 31-40 GeV DM particles distributed according to a (contracted) NFW profile and annihilating into $b\bar{b}$ with $\langle \sigma v\rangle = 1.4 \div 2 \times 10^{-26}$ cm$^3/{\rm s}$. These results have understandably spurred an intense model building activity from the community~\cite{hooperon model building}, possibly because the main features of this potential signal are very close to widespread expectations (GC origin, close-to-thermal annihilation cross-section, `conventional' mass and annihilation channel, matching to one of the most popular DM distribution profiles...). Some models may also provide an interesting connection with other recent astroparticle anomalies such as the {\sc Dama/Libra} and {\sc Integral} signals~\cite{hooperon spinoffs}. In the light of these points, it is interesting and timely to assess the robustness of the DM claims against associated constraints, as we set out to do.

\medskip

Before moving on, however, two important cautionary remarks are in order. Firstly, one should not forget that, in very general terms, the identification of an `excess' strongly relies on the capability of carefully assessing the background over which the excess is supposed to emerge. As we already implicitly suggested above, the claim under scrutiny constitutes no exception, quite the contrary. The extraction of the residuals strongly relies on the modeling of the diffuse gamma-ray background (in particular the one publicly made available by the {\sc Fermi} collaboration) as well as on additional modeling of astrophysical emissions, e.g. from {\sc Fermi} bubbles, isotropic component, unresolved point sources, molecular gas... 
While this is probably the best that can be done, it is not guaranteed to be (and in general is not expected to be) the optimal strategy. 

Secondly, one should not forget that there might be alternative astrophysical explanations for the excess. A population of milli-second pulsars has been extensively discussed since the beginning~\cite{MSP}, as well as the possibility of a spectral break in the emission of the central Black Hole~\cite{Boyarsky:2010dr}. More recently, the possibility has been suggested that isolated injections of charged particles (electrons~\cite{Petrovic:2014uda} or protons~\cite{Carlson:2014cwa}) sometime in the past, possibly connected with the activity of the central Black Hole, can produce secondary radiation able to account for the anomalous signal. While reproducing with these models all the details of the observed emission might be not easy, they represent plausible and useful counterexamples to the DM interpretation. 

\medskip

Keeping these points in mind, we wish to insist on the tantalizing DM hypothesis and we wish to explore ways to confirm or disprove the result within the DM framework. To this purpose, it is useful to follow a multi-messenger approach. In particular, given the alleged hadronic origin of the signal, it is very useful to analyze the antiproton channel to put constraints on the DM interpretation of such excess. The antiprotons are a precious cross check in this context for several reasons: first of all, it is common wisdom that --in most non-leptophilic WIMP scenarios-- the ratio between the exotic signal and the astrophysical background is noticeably large in this channel; moreover, the astrophysical background is known with a considerable level of precision: it depends on the proton flux and, apart from an unavoidable uncertainty on the production cross-section, it is not highly affected by the choice of a diffusion setup among those allowed by light nuclei ratios and other observables. In the light of this, it is not surprising that antiproton measurements have been providing since a long time a powerful probe for imposing constraints on any exotic signal, notably due to DM~\cite{pbar history}. On the other hand, the signal coming from a light DM particle is expected to lie in an energy range where the effect of solar modulation is very important, making the task of identifying an extra signal much more complicated.

Very recently, ref.~\cite{Bringmann:2014lpa} performed an analysis of antiprotons constraints which partially overlaps with our scope. We will compare in detail our respective results later on; now we can anticipate that we adopt a different strategy concerning solar modulation, we use different tools (in particular we adopt numerical packages such as {\tt DRAGON} and {\tt GammaSky}), and, overall, we find less stringent and more conservative bounds.
 
\medskip

Another key issue in the analysis of the GC residuals is the role of the secondary gamma radiation (via Bremsstrahlung and Inverse Compton processes, respectively on the Galactic gas and ambient light) emitted from the particles originating from DM annihilation. Already in~\cite{Cirelli:2013mqa} it was pointed out that, for the ranges of energies under discussion and for the gas-rich regions close to the GC, Bremsstrahlung is the dominant process of energy loss for the electrons and positrons produced by DM annihilation, resulting in an important contribution to the gamma ray emission. Ref.~\cite{Lacroix:2014eea} has analyzed the issue in further detail, showing that leptonic channels can actually provide a better fit of the GC excess when the secondary emissions are included. However, that study still implements an approximated framework for energy losses. In the analysis of~\cite{Abazajian:2014fta} and~\cite{Daylan:2014rsa}, the impact of Bremsstrahlung is discussed too, although only at the level of an estimate. Here we go one step further: we employ {\tt DRAGON}~\cite{evoli:2008,gaggero:2013} and {\tt GammaSky}~\cite{GammaSky} in order to compute realistic and accurate gamma-ray spectra including all secondary radiations for annihilations into a couple of exemplar leptonic channels. We then compare the spectra with data, as extracted from~\cite{Daylan:2014rsa} and determine the best fit regions. This is not intended as a fully thorough analysis (in particular because the data are extracted using templates that do not account for secondary emissions), but it provides a useful example of how important a proper calculation of the full DM gamma-ray emission can be. Moreover, in the light of the antiprotons constraints disfavoring the hadronic channels, it will be important to assess the viability of the alternative leptonic channels. 

\medskip

The rest of this paper is organized as follows. 
In section~\ref{sec:preliminaries} we briefly recall the basics of the computation of the gamma ray spectra from DM in the GC and of charged particles in the Galaxy, and we introduce the tools we will use in the following.
In section~\ref{sec:fits} we present our determination of the exemplary best fit regions for the $b\bar b$ and leptonic channels.  
In section~\ref{sec:antiproton bounds} we compute the bounds from the antiproton measurements, discussing in particular the impact of solar modulation.
Finally in section~\ref{sec:conclusions} we summarize and conclude.

%%%%%%%%%%%%%%%%%%%%%%%%%%%%%%%%%%%%%%%%%%%%%%%%%%%%%%%%%
%%%%%%%%%%%%%%%%%%%%%%%%%%%%%%%%%%%%%%%%%%%%%%%%%%%%%%%%%

\section{Dark Matter and Cosmic Rays: setups and tools}
\label{sec:preliminaries}

In this section we briefly present the basic ingredients that we will need for the subsequent computations, both in terms of physics assumptions and in terms of technical tools.

\subsection{DM Galactic distribution and DM gamma-ray flux}
\label{sec:DMdistribution}

In general, the DM distribution $\rho$ in the Galaxy is of course unknown and it is actually source of significant uncertainty. 
In this specific case, however, it is claimed that the morphology of the GC excess signal allows to determine $\rho$ quite precisely. The best choice, which we also adopt, appears to be a generalized Navarro-Frenk-White profile (gNFW)~\cite{gNFW} defined as
\begin{equation}\label{eq:gNFW}
\rho_{\rm gNFW}(r) = \rho_{\odot}\left(\frac{r}{r_{\odot}}\right)^{-\gamma}
\left[
\frac{1+(r/R_s)^{\alpha}}{1+(r_{\odot}/R_s)^{\alpha}}
\right]^{-\frac{(\beta-\gamma)}{\alpha}}~,
\end{equation}
with $\rho_{\odot} = 0.3$ GeV/cm$^{3}$, $r_{\odot} = 8.5$ kpc, $R_s=20$ kpc, $\alpha = 1$, $\beta = 3$. The parameter $\gamma$ controls the inner slope of the profile. For the standard NFW profile one has $\gamma = 1.0$. The GC excess is better fit by a profile with $\gamma = 1.20$ or 1.26~\cite{Daylan:2014rsa}, corresponding to an inner portion steeper than the standard one (i.e.~`contracted'), and therefore we will use these two values in the following.

\medskip

The prompt gamma-ray differential flux from DM annihilation from a given angular direction $d\Omega$ is given by
\begin{equation}
\frac{d\Phi}{dE_{\gamma}d\Omega} = \frac{r_{\odot}}{8\pi}
\left(\frac{\rho_\odot}{M_{\rm DM}}\right)^{2}~J~\sum_f \langle \sigma v\rangle_f \frac{dN_{\gamma}^f}{dE_{\gamma}}~,
\end{equation}
with 
\begin{equation}
J \equiv \int_{\rm l.o.s.}\frac{ds}{r_{\odot}}\left[\frac{\rho_{\rm gNFW}(r(s,\theta))}
{\rho_{\odot}}\right]^2~,
\end{equation}
and $r(s,\theta) = (s^2 + r_{\odot}^2 - 2 sr_{\odot}\cos\theta)^{1/2}$, $\theta$ being the aperture angle between the direction of the line of sight and the axis connecting the Earth to the GC. We take the input spectra $dN_{\gamma}^f/dE_{\gamma}$ for a final state $f$ (for instance $b \bar b$) from the {\tt PPPC4DMID}~\cite{PPPC4DMID}. They include ElectroWeak radiations and other refinements with respect to previous computations, although these are mostly not relevant in our case as we are interested in the low DM mass case.

DM also emits secondary gamma-rays, namely via the already mentioned IC and Brems processes. In the former, the electrons and positrons originating from the annihilation hit an ambient photon (from the CMB, the InfraRed or the Optical galactic radiation field) and up-scatter it to gamma-ray energies. In the latter, those same electrons suffer bremsstrahlung on the (neutral or ionized) interstellar gas and radiate soft gamma rays. 
In analogy with the prompt one, the IC gamma-ray differential flux can be conveniently expressed as~\cite{PPPC4DMID}
\begin{equation}
\frac{d\Phi_{{\rm IC}\gamma}}{dE_\gamma\, d\Omega} = \frac1{E_\gamma^2}\frac{r_\odot}{4\pi}  \frac12 \left(\frac{\rho_\odot}{M_{\rm DM}}\right)^2 \int_{m_e}^{M_{\rm DM}} \hspace{-0.45cm} dE_{\rm s}  \sum_f \langle \sigma v \rangle_f \frac{dN^f_{e^\pm}}{dE}(E_{\rm s})  \, I_{\rm IC}(E_\gamma,E_{\rm s},b,\ell)
\label{eq:summaryIC}
\end{equation}
where the IC halo functions $I_{\rm IC}$ incorporate all the details of the energy loss and IC processes. A completely analogous formalism can be worked out for Brem emission. We refer to~\cite{PPPC4DMID} and~\cite{PPPC4DMsyn} for further details and for an updated version.  

\medskip

In practice, however, we compute the IC and Brem $\gamma$-ray fluxes using {\tt GammaSky}. This numerical package is interfaced to {\tt DRAGON} (see the next paragraph for a description of {\tt DRAGON}) and computes the line of sight integral of the the gamma-ray, synchrotron and neutrino emissions originating from the interactions of the CRs (output by {\tt DRAGON}) with the interstellar gas, the magnetic field and the diffuse radiation field (from microwaves to UV). 
Some realistic magnetic field distributions (as described in \cite{GammaSky}), gas and interstellar radiation field (ISRF) models  are implemented. In particular, the gas and ISRF we use here are the same as in the latest public version of {\tt GalProp}~\cite{Galprop}. One should recall that the gas distributions have large uncertainties and are known to map only approximatively the regions close to the GC (see e.g. the discussion in~\cite{Cirelli:2013mqa}), although in practice this will have a limited impact on our work as we consider areas far enough from the GC.

The code will be published soon; some results obtained with this tool can be found in~\cite{GammaSky}.
For this work we have also cross checked the output of the code, for specific cases, against the semi-analytic calculations mentioned above, finding a very good agreement.

\subsection{Charged cosmic rays propagation}
\label{sec:CR}

For propagating charged cosmic rays in the Galaxy (electrons, positrons, protons and antiprotons), we use {\tt DRAGON}~\cite{evoli:2008,gaggero:2013}. 
This numerical package solves the general diffusion-loss equation:
\begin{eqnarray}
\label{eq:diffusion_equation}
\frac{\partial N^i(r, z, p)}{\partial t} &=&  \fraction{\partial}{\partial x_i} D_{ij} \fraction{\partial N^{i}}{\partial x_j} 
+\bm{v}_{c}\cdot{\bm \nabla} N^{i}  \\
&-& \frac{\partial}{\partial p} \left(\dot{p}-\frac{p}{3}\bm{\nabla}\cdot\bm{v}_{c}\right) N^i +\frac{\partial}{\partial p} p^2 D_{pp}
\frac{\partial}{\partial p} \frac{N^i}{p^2} +   Q(r,z,p)\nonumber
\end{eqnarray} 
for any relevant species $N^i$, from heavier nuclei down to protons, antiprotons and leptons, coming from both astrophysical processes and DM annihilation or decay (as encoded in the source term $Q$). The code can perform computations in 2D, 3D isotropic and 3D anisotropic mode; since the impact of three-dimensional structures in the Galaxy is not relevant for the antiproton channel in which we will be mainly interested, we will work in 2D mode. 
All the terms in eq.~(\ref{eq:diffusion_equation}) are described in detail in the literature (e.g.~\cite{evoli:2008} and references therein) and we do not repeat them here.
We just specify that, for the diffusion term (the first term on the r.h.s), we adopt a scalar diffusion coefficient with the following dependence on the particle rigidity ($R$) and on the distance from the Galactic Plane ($z$):
\begin{equation}
D \,\, = \,\, D_0 \, \beta^{\eta} \, \left({\frac{R}{R_0}}\right)^{\delta} \, e^{|z|/z_t}.
\end{equation}
In this equation several free parameters appear:
\begin{itemize} 
\item[-] $D_0$ is the normalization of the diffusion coefficient at the reference rigidity $R_0$ = 3 GeV;
\item[-] $\eta$ parametrizes our poor knowledge of the complicated physical effects that may play a major role at low energy (below 1 GeV), e.g. the dissipation of Alfv\'en waves due to the resonant interaction with CRs (here $\beta$ represents as usual the particle velocity in units of the speed of light);
\item[-] $\delta$ is the slope of the power law in rigidity and is constrained by light nuclei ratios, in particular Boron-over-Carbon (B/C);
\item[-] $z_t$ is the scale height of the diffusive halo of the Galaxy. As shown in \cite{evoli:2008}, the exponentially growing behaviour that we assume is physically more realistic than a uniform one and allows to get a more regular CR density profile at the vertical boundaries of the halo.
\end{itemize}
For the DM source, we feed {\tt DRAGON} with the input spectra from {\tt PPPC4DMID}~\cite{PPPC4DMID}. 

\medskip

In the final stage of the propagation process, charged particles have to penetrate into the sphere of influence of the Sun and are subject to the solar modulation effect, very relevant for low energy ($\lesssim$ 10 GeV) particles. In general terms, the solar wind decreases the kinetic energy $K$ and momentum $p$ of charged cosmic rays. 
This can be effectively described in the so-called `force field approximation': specifying the notation to the case of antiprotons, the energy spectrum $d\Phi_{\bar{p}\oplus}/dK_\oplus$
of particles that reach the Earth with energy $K_\oplus$ and momentum $p_\oplus$ is approximatively related to their energy spectrum in the interstellar medium, $d\Phi_{\bar p}/dK$,
as~\cite{GA}
\begin{equation}\label{eq:Fisk}
\frac{d\Phi_{{\bar p}\oplus}}{dK_\oplus} = \frac{p_\oplus^2}{p^2} \frac{d\Phi_{\bar p}}{dK},\qquad
K= K_\oplus + |Ze| \phi_F^{\bar p}, \qquad
p^2 = 2m_p K+K^2.
\end{equation}
The so-called Fisk potential $\phi_F^{\bar p}$ parameterizes in this effective formalism the kinetic energy loss.

\medskip
     
However, since the modulation process -- due to relevant drift effects -- depends on the charge of the particle (including its sign), it is useful in many cases to go beyond the very simplified picture of the force field.
A more accurate treatment consists in using a dedicated numerical transport tool for this part of the propagation process, namely {\tt HelioProp}~\cite{helioprop}. 
This code computes the propagation of CRs in the heliosphere and implements a realistic model for the structure of the solar magnetic field (SMF). 
The SMF changes its properties during the 22-year solar cycle and is modeled in this package by a Parker spiral, with opposite polarities in the northern and southern hemispheres; the presence of a heliospheric current sheet at the interface between the regions with opposite polarities is taken into account.
In general it is found that, especially near the minimum of the 22-year solar cycle, the force field approximation can reproduce well the detailed results of the dedicated transport method. 
We will hence make use of the results of {\tt HelioProp} and transpose them into a determination of the range of $\phi_F$ to be plugged in the context of the force field approximation (see later for more details).

\medskip

\begin{table}[!t]
\centering
\begin{tabular}{|c|c|c|c|c|c||c|c|}\hline
      & {\color{magenta}{KRA}} & {\color{blue}{KOL}} & {\color{brown}{CON}} & {\color{orange}{THK}} &  {\color{forestgreen}{THN}} &   {\color{forestgreen}{THN2}} & {\color{forestgreen}{THN3}}   \\ \hline
       $z_t$ [kpc]    &  4 & 4 & 4 & 10 & 0.5 & 2 & 3 \\ \hline
           $D_{0}$ [$10^{28}$ cm$^2$\,s$^{-1}$]  & 2.64 & 4.46 & 0.97 & 4.75 & 0.31 & 1.35 & 1.98 \\ \hline
       $\delta$  & 0.50 & 0.33 & 0.6 & 0.50 & 0.50 & 0.50 & 0.50         \\ \hline
       $\eta$ & -0.39 & 1 & 1 & -0.15 & -0.27 & -0.27 & -0.27    \\ \hline
        $v_{\rm A}$ [km\,s$^{-1}$] & 14.2 & 36 & 38.1 & 14.1 & 11.6  & 11.6 & 11.6       \\ \hline
        $\gamma$  & 2.35 & 1.78/2.45 & 1.62/2.35 & 2.35 & 2.35 & 2.35 & 2.35       \\ \hline
              $dv_{\rm c}/dz$\,[ km\,s$^{-1}$\,kpc$^{-1}$]& 0 & 0 & 50 & 0 &  0 & 0 & 0       \\ \hline\hline
          $\phi_F^p$ [GV]    &  0.650 & 0.335 & 0.282 & 0.687 & 0.704 & 0.626 & 0.623 \\ \hline
                  $\chi^2_{\rm min}/{\rm dof}$ ($p$ in~\cite{Adriani:2011cu})  & 0.462 & 0.761 & 1.602 & 0.516 & 0.639 &  0.343 & 0.339  \\ \hline
                        \end{tabular}\vspace{0.3cm}
                        \caption{\label{tab:Phi} \textit{
                        Values of the propagation parameters for the 5 standard profiles of ref.~\cite{Evoli:2011id} (left portion of the table) as well as of the modified `thin' setups that we will use in the following (right portion). For each setup we report also the best-fit value of the solar modulation potential for protons $\phi_F^p$   obtained against the proton data in ref.~\cite{Adriani:2011cu}.          }}
     \end{table}

\begin{figure}[h]
\begin{center}
\centering
  \begin{minipage}{0.49\textwidth}
   \centering
   \includegraphics[scale=0.45]{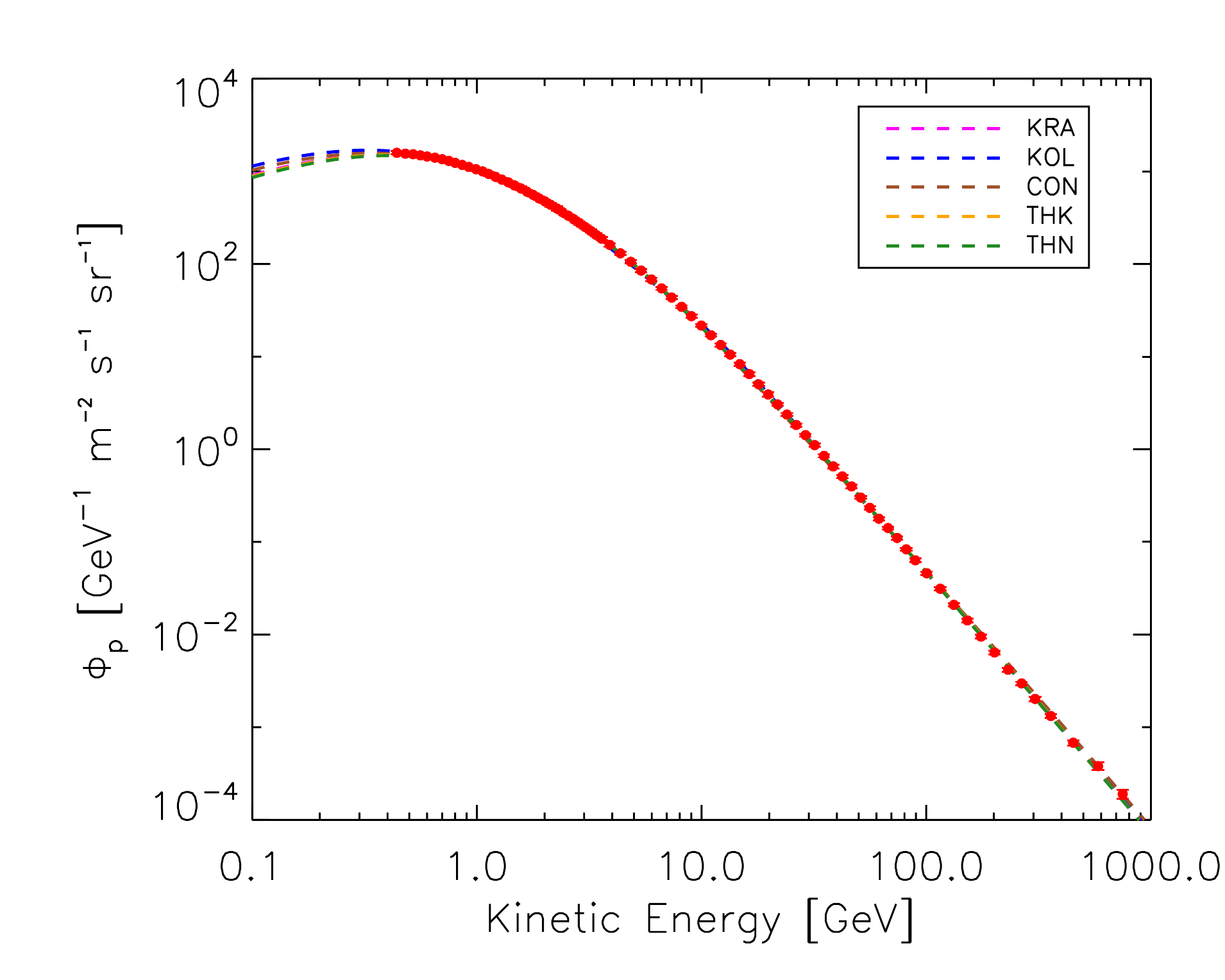}
   %\caption{\textit{Count Map}}\label{fig:CountMap}
    \end{minipage}\hspace{0.05 cm}
   \begin{minipage}{0.49\textwidth}
    \centering
    \includegraphics[scale=0.45]{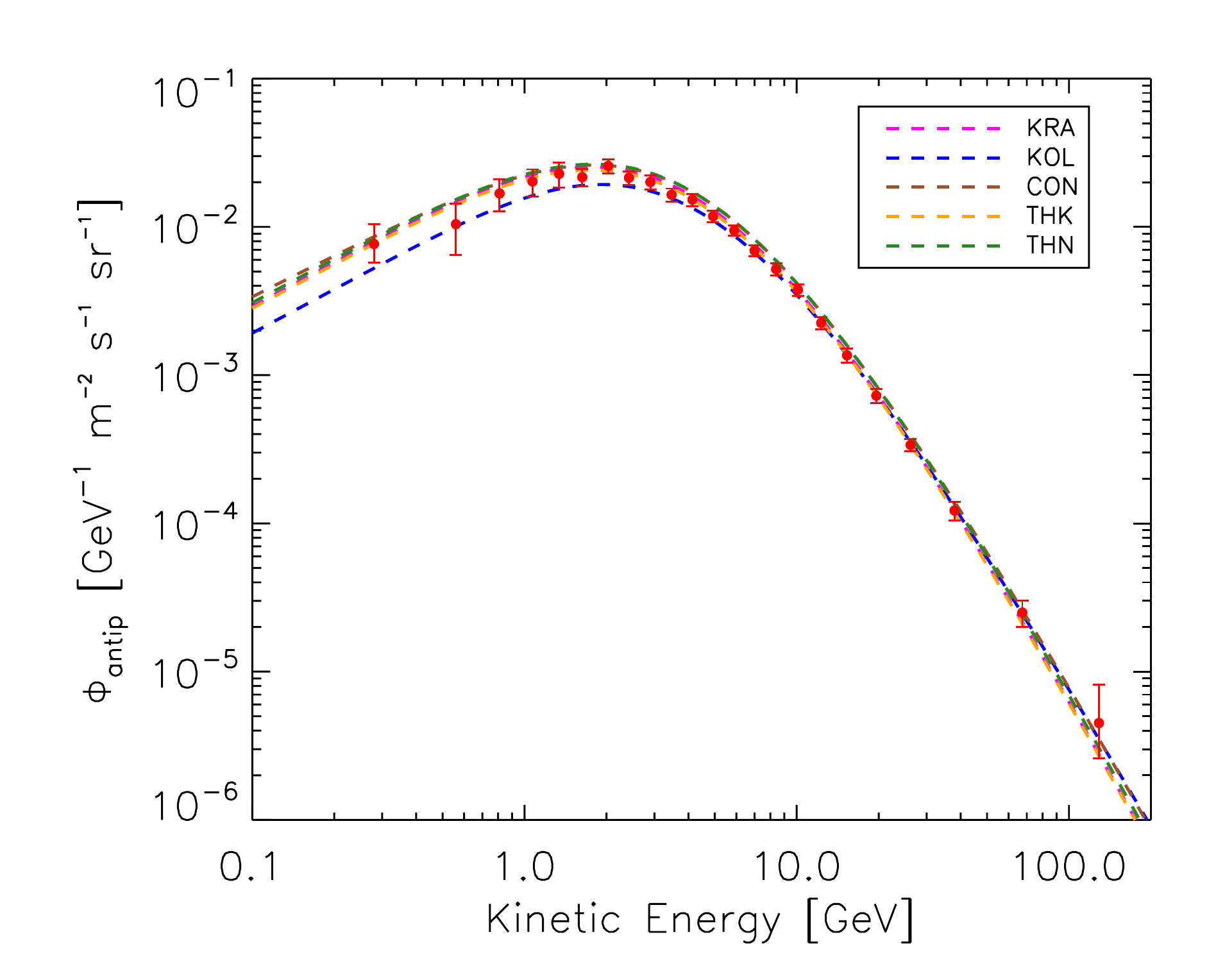}
    \end{minipage}\\
\caption{\textit{
Cosmic-ray proton (left panel) and antiproton (right panel) flux measured by the {\sc Pamela} experiment.  We superimpose to the experimental data -- taken, respectively, from ref.~\cite{Adriani:2011cu} and ref.~\cite{Adriani:2012paa} -- the background estimations obtained using the five propagation models defined in ref.~\cite{Evoli:2011id}. Fitting the proton data we determine the best solar modulation potential $\phi_F^p$ for each setup (we give the corresponding values in table~\ref{tab:Phi}). For the purpose of this figure we modulate the $\bar p$ flux with the same value.
}}
\label{fig:DataFit}
\end{center}
\end{figure}

The free parameters that appear in the diffusion-loss equation (\ref{eq:diffusion_equation}) define a {\it propagation setup}. 
%The setup can be constrained by secondary-to-primary ratios such as B/C and other observables (protons, beryllium, other ratios such as N/O and subFe/Fe...); 
As discussed e.g.~in~\cite{Evoli:2011id}, the uncertainty on the diffusion parameters produce a modest spread in predictions for the antiproton flux coming from $p$-$p$ and $p$-He collisions, while the impact on the flux coming from DM annihilations is much larger.

We adopt the five propagation models defined in ref.~\cite{Evoli:2011id}. We summarize their properties in table~\ref{tab:Phi} and comment on them as follows.
\begin{itemize}
\item[$\circ$] The THN, KRA and THK models assume the same value of $\delta$ -- corresponding to a Kraichnan-type turbulence in the Quasi-Linear Theory (QLT) -- but different values for the height $z_t$ of the diffusion cylinder: THN corresponds to a very thin diffusion zone (0.5 kpc), KRA assumes 4 kpc and THK applies if the diffusion zone is as thick as 10 kpc.
\item[$\circ$] The KOL model instead assumes a $\delta = 0.33$ (which is given by Kolmogorov turbulence in QLT), with a diffusive characteristic height fixed at 4 kpc. 
\item[$\circ$] The CON model includes strong convective effects (but the diffusive height is still fixed at 4 kpc).
\end{itemize}
As pointed out e.g.~in~\cite{Evoli:2011id} (among others), the most relevant uncertainty for this kind of analysis is the thickness of the diffusion zone $z_t$: a thinner halo corresponds to a much lower signal for the antiprotons coming from DM annihilation, hence one anticipates that the constraints obtained using this setup will be much weaker.
In order to investigate more carefully different choices for the thickness of the diffusion zone, we define two more THN-type models with, respectively, $z_t=2,\,3$ kpc (see table~\ref{tab:Phi}). 

For each propagation setup we fit  the solar modulation potential $\phi_F^p$ against the {\sc Pamela} proton data in ref.~\cite{Adriani:2011cu} (we report the resulting values in the table~\ref{tab:Phi}). We show the result of this analysis in fig.~\ref{fig:DataFit}. 
Concerning the Fisk potential for antiprotons (a crucial quantity, as we will see in the following) we will adopt several strategies, discussed in section~\ref{sec:antiproton bounds}.

Before moving on, let us stress one important limit of this description. The diffusion-loss equation~\ref{eq:diffusion_equation} is solved in the steady-state approximation, 
i.e. $\partial N^i(r, z, p)/\partial t  = 0$. The possibility to go beyond this approximation is, at the moment, a major and only partially addressed task that needs to be improved in the future in order to have a better control on the propagation of cosmic rays in the Galaxy.

%%%%%%%%%%%%%%%%%%%%%%%%%%%%%%%%%%%%%%%%%%%%%%%%%%%%%%%%%
%%%%%%%%%%%%%%%%%%%%%%%%%%%%%%%%%%%%%%%%%%%%%%%%%%%%%%%%%

\section{Fits of the gamma-rays from DM annihilation in the GC}
\label{sec:fits}

We focus on two benchmark cases: 100\% DM annihilation into $b\bar{b}$ final state and annihilation into leptons in some specific mixtures of flavors as indicated in Table~\ref{tab:result}. The latter ones are chosen since, as we will see, they are very close to producing the best fit to the data with which we will be comparing. But since such data depend on the details of the analysis and might change as more refined background subtractions are developed, they can just be considered as typical examples for a leptonic channel.

Following the results of ref.~\cite{Daylan:2014rsa}, we analyze two different sets of residual data describing the gamma-ray emission associated with DM.
In the `Galactic Center' analysis the region of interest is defined as the region $|b|<5^{\circ}$, $|l|<5^{\circ}$, while the `Inner Galaxy' analysis is based on a full-sky fit (masking $1^{\circ}$  in latitude around the galactic plane). In the first setup the best-fit value for the slope of the gNFW profile in eq.~(\ref{eq:gNFW}) turns out to be $\gamma = 1.2$, while the second approach seems to prefer a slightly larger value, $\gamma = 1.26$. In both cases we compute -- using the corresponding values of $\gamma$ -- the differential flux from DM annihilation considering one specific l.o.s. with $\theta = 5^{\circ}$ and compare with the data presented in ref.~\cite{Daylan:2014rsa}, which are normalized under this assumption~\cite{Slatyer}.

\begin{figure}[!t]
\centering
 \begin{minipage}{\textwidth}
   \centering
  \includegraphics[width=1 \linewidth]{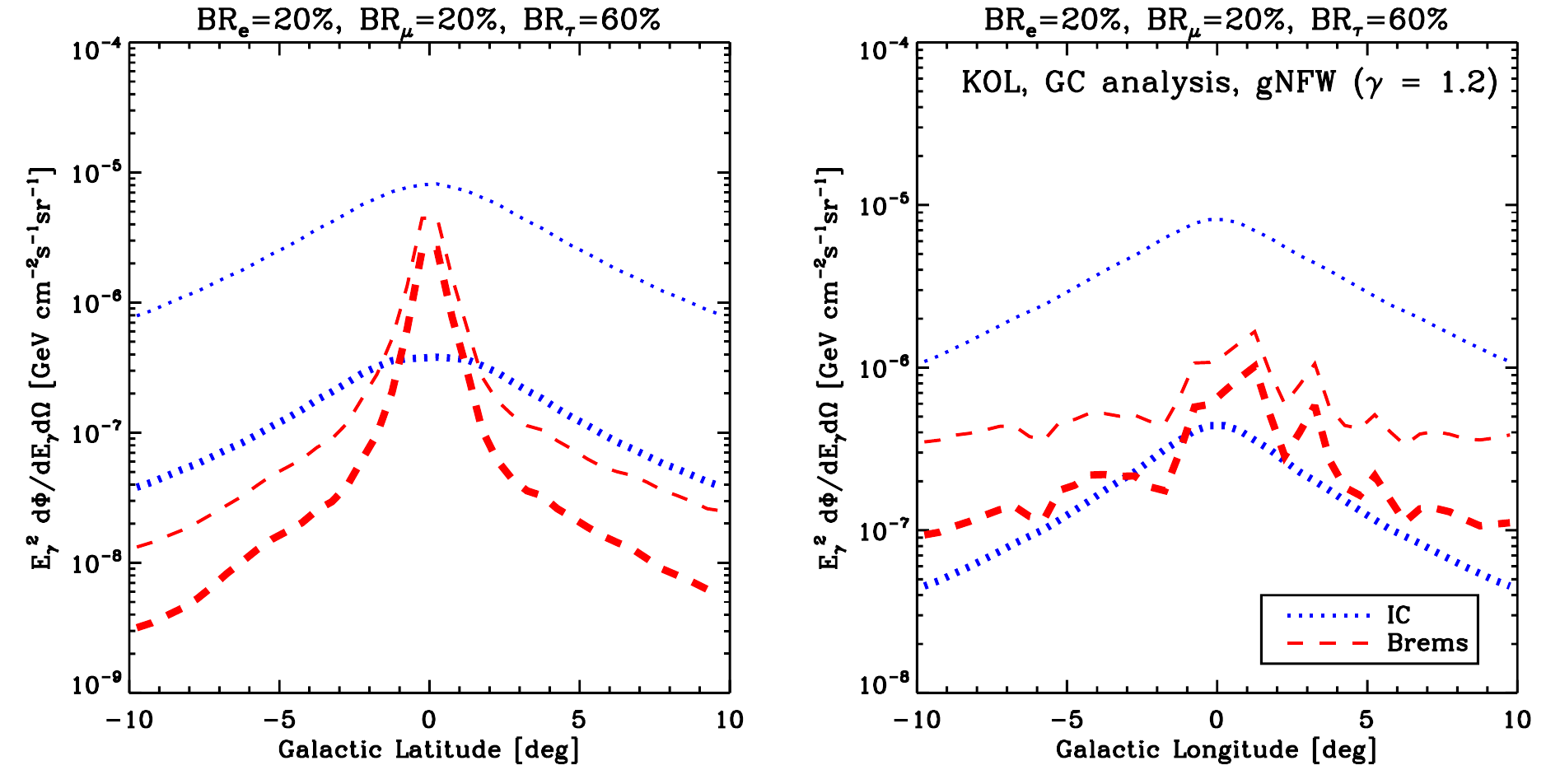}
   \end{minipage}\vspace{1 cm}\\
     \begin{minipage}{0.49\textwidth}
   \centering
   \includegraphics[scale=0.32]{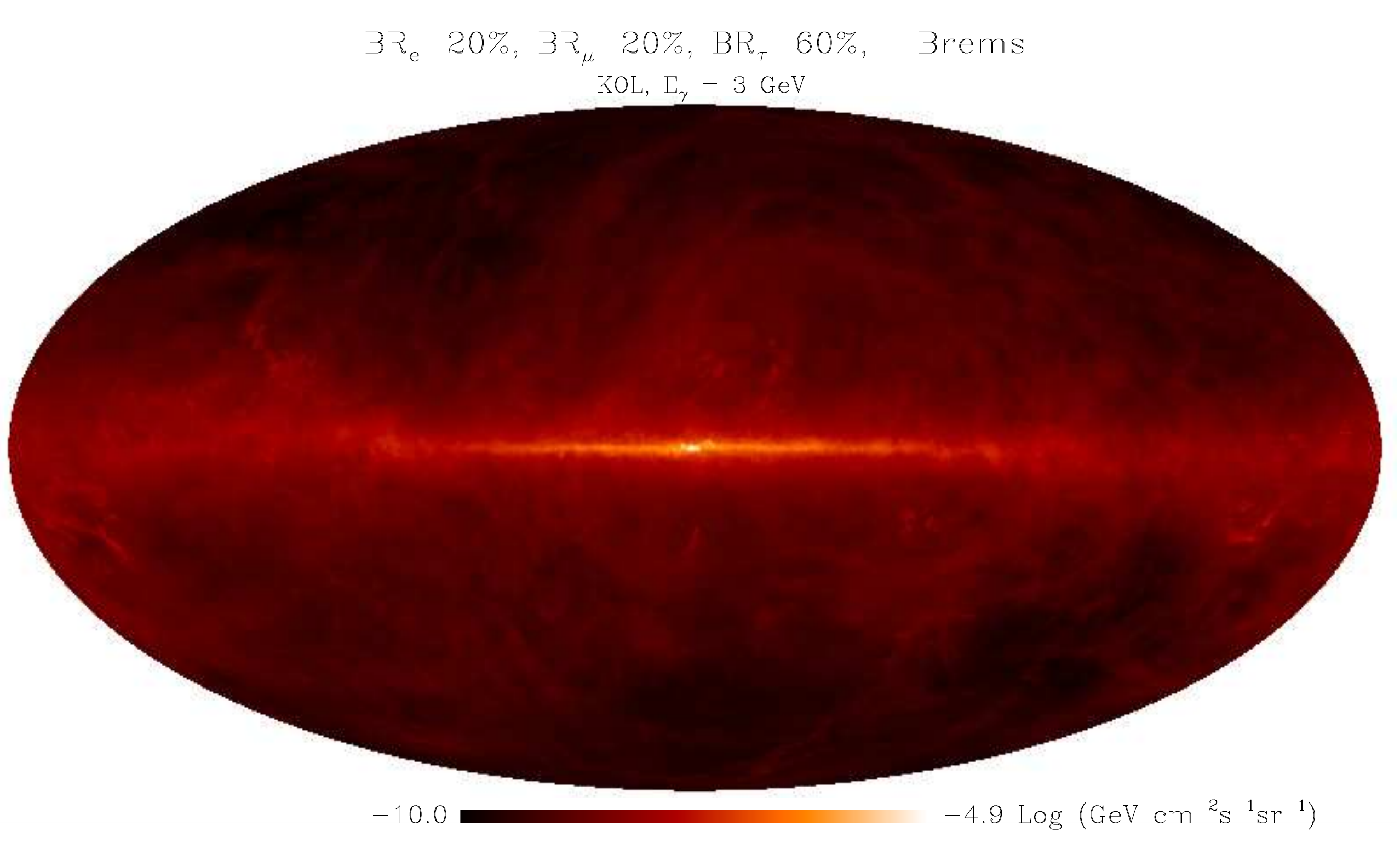}
   %\caption{\textit{Count Map}}\label{fig:CountMap}
    \end{minipage}\hspace{0.1 cm}
   \begin{minipage}{0.49\textwidth}
    \centering
    \includegraphics[scale=0.32]{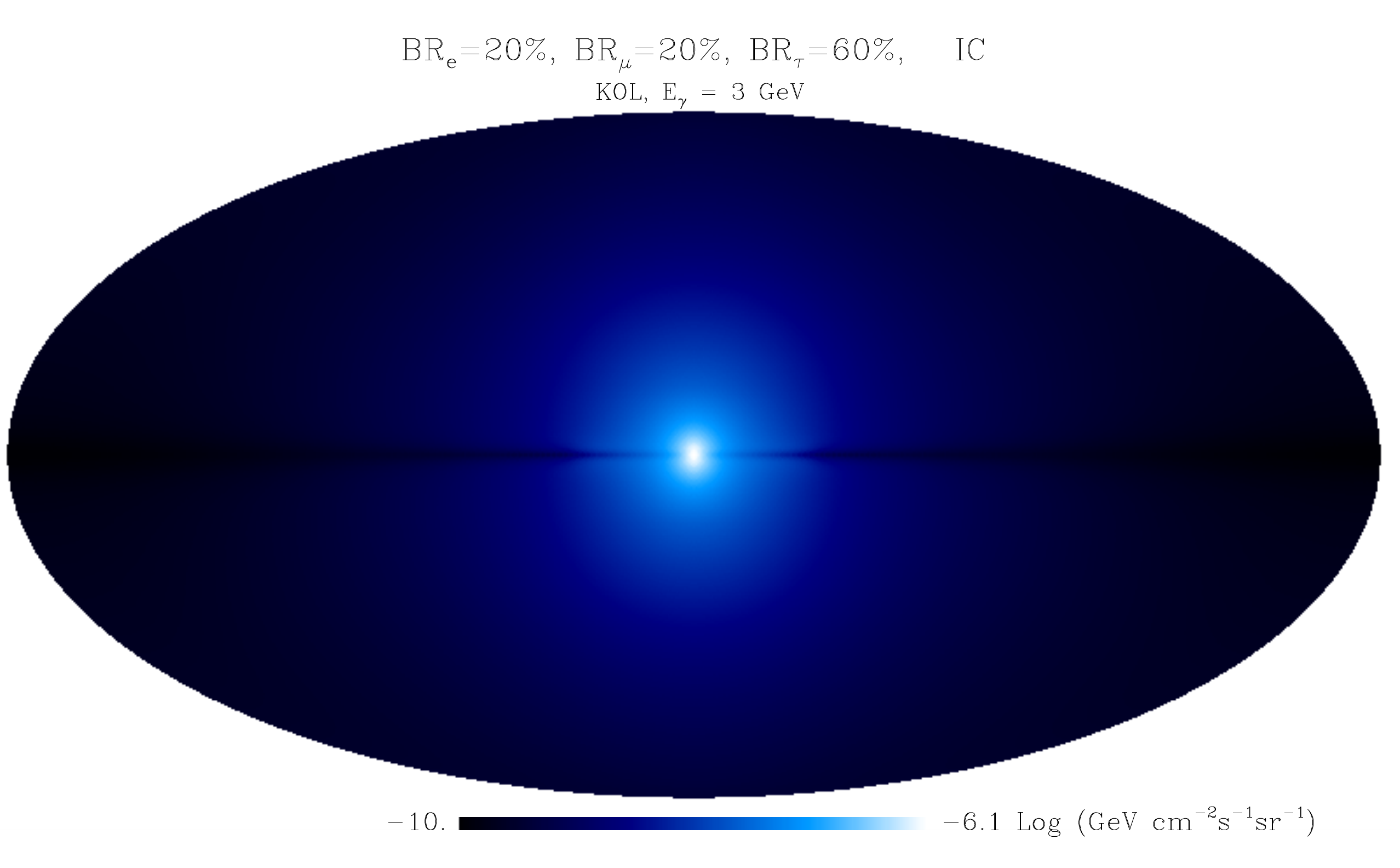}
    \end{minipage}
 \caption{\textit{Upper panels:
 latitude- and longitude-dependence of Brems and IC emissions from DM annihilation 
 into leptons with $\langle \sigma v \rangle =3\times 10^{-26}$ cm$^3$\,s$^{-1}$ and branching ratios as given in table~\ref{tab:result}. We show secondary emissions for $E_{\gamma} = 3$ GeV (thicker lines) and $E_{\gamma} = 0.5$ GeV (thinner lines). Lower panels: representative sky maps for Brems (left) and IC (right) emission, for the same choices of parameters.
 }}\label{fig:Morphology}
\end{figure}

A comment concerning the morphology of the different emissions is now in order. The prompt gamma-ray flux is of course spherically symmetric. For the secondary emissions, in order to have a better insight on their morphology we plot in fig.~\ref{fig:Morphology} --considering the DM annihilation into leptons, where the impact of this kind of emission is larger-- the Brems and IC emission as a function of the Galactic latitude (upper-left panel, averaging on $|l| \leqslant 5^{\circ}$) and longitude (upper-right panel, averaging on $|b| \leqslant 5^{\circ}$) for two representative values of the energy, $E_{\gamma}=0.5$ GeV (thinner lines) and $E_{\gamma}=3$ GeV (thicker lines). For completeness, we also show in the lower panel of fig.~\ref{fig:Morphology} the corresponding sky maps for $E_{\gamma}=3$ GeV.
\begin{figure}[!htb!]
\centering
  \includegraphics[width=1 \linewidth]{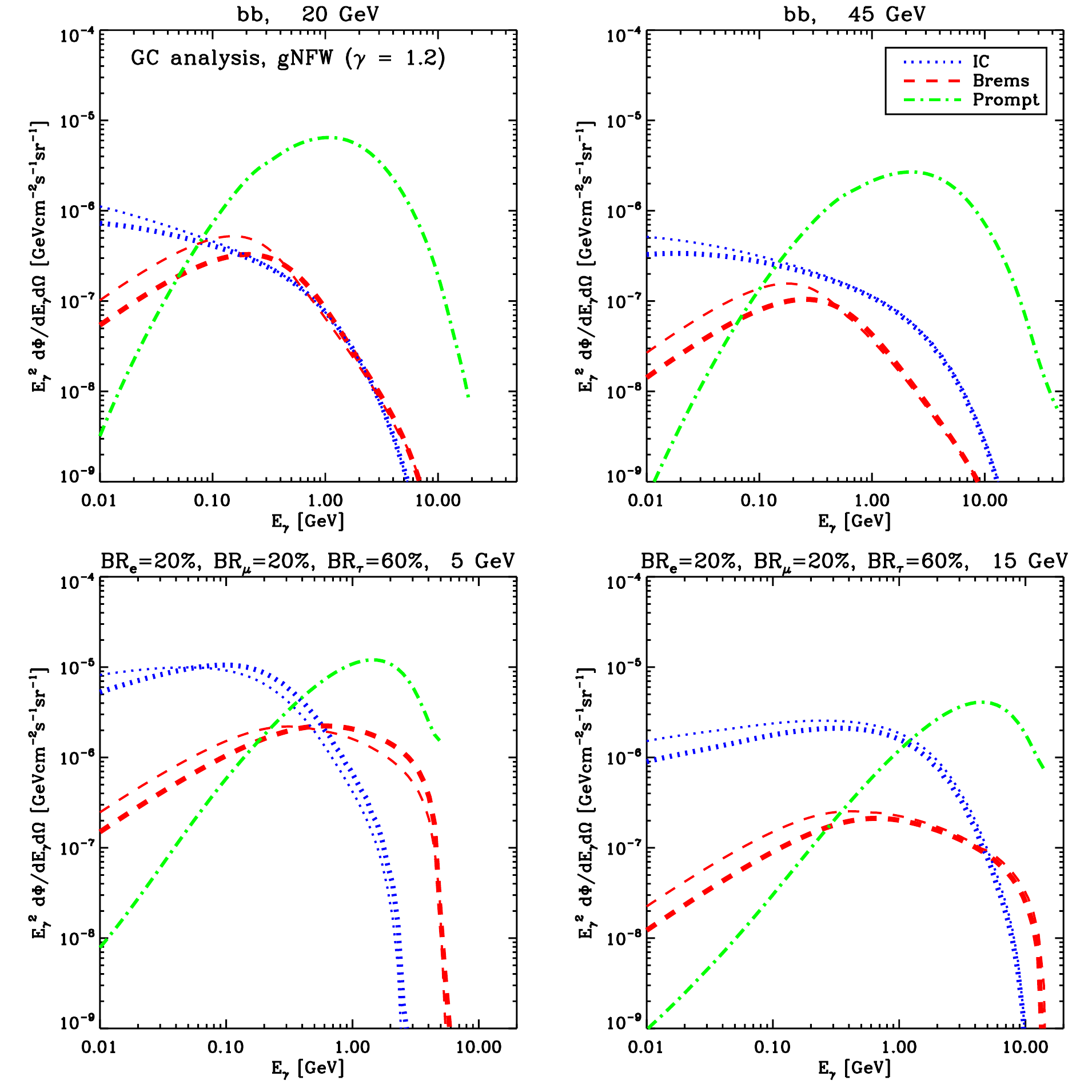}
 \caption{\textit{
 Gamma-ray spectrum from DM annihilation in the GC. We consider 100\% DM annihilation into $b\bar{b}$ final state (upper panels) and DM annihilation into leptons with branching ratios as given in table~\ref{tab:result} (lower panels). We separately show prompt, IC and Brems emissions for four different representative values of the DM mass. We assume a gNFW profile with $\gamma = 1.2$, and we take $\langle \sigma v\rangle  = 3\times 10^{-26}$ cm$^3$\,s$^{-1}$. Thicker (thinner) lines correspond to the KOL (THN) propagation model. 
 }}\label{fig:BBSpectra}
\end{figure}
This allows to see that, while the IC emission is to a good approximation spherically symmetric too, the Brems emission, which is correlated with the gas density in the Galactic Plane, is far from being so, not surprisingly. In order to meaningfully compare with the data in~\cite{Daylan:2014rsa}, derived under the assumption of spherical symmetry, we take the averaged value of the differential flux in a Galactocentric disk with $\theta \in [4.8^{\circ}, 5.2^{\circ}]$. A more accurate treatment, which is however out of the scope of our current analysis, would consist in extracting the data by including secondary emission, especially Brems, in the DM template used to fit the residual. 
%{\color{red} TBC} We leave this kind of analysis for future work. 

\medskip

\begin{figure}[!t]
\begin{center}
\centering
  \begin{minipage}{0.45\textwidth}
   \centering
   \includegraphics[scale=0.61]{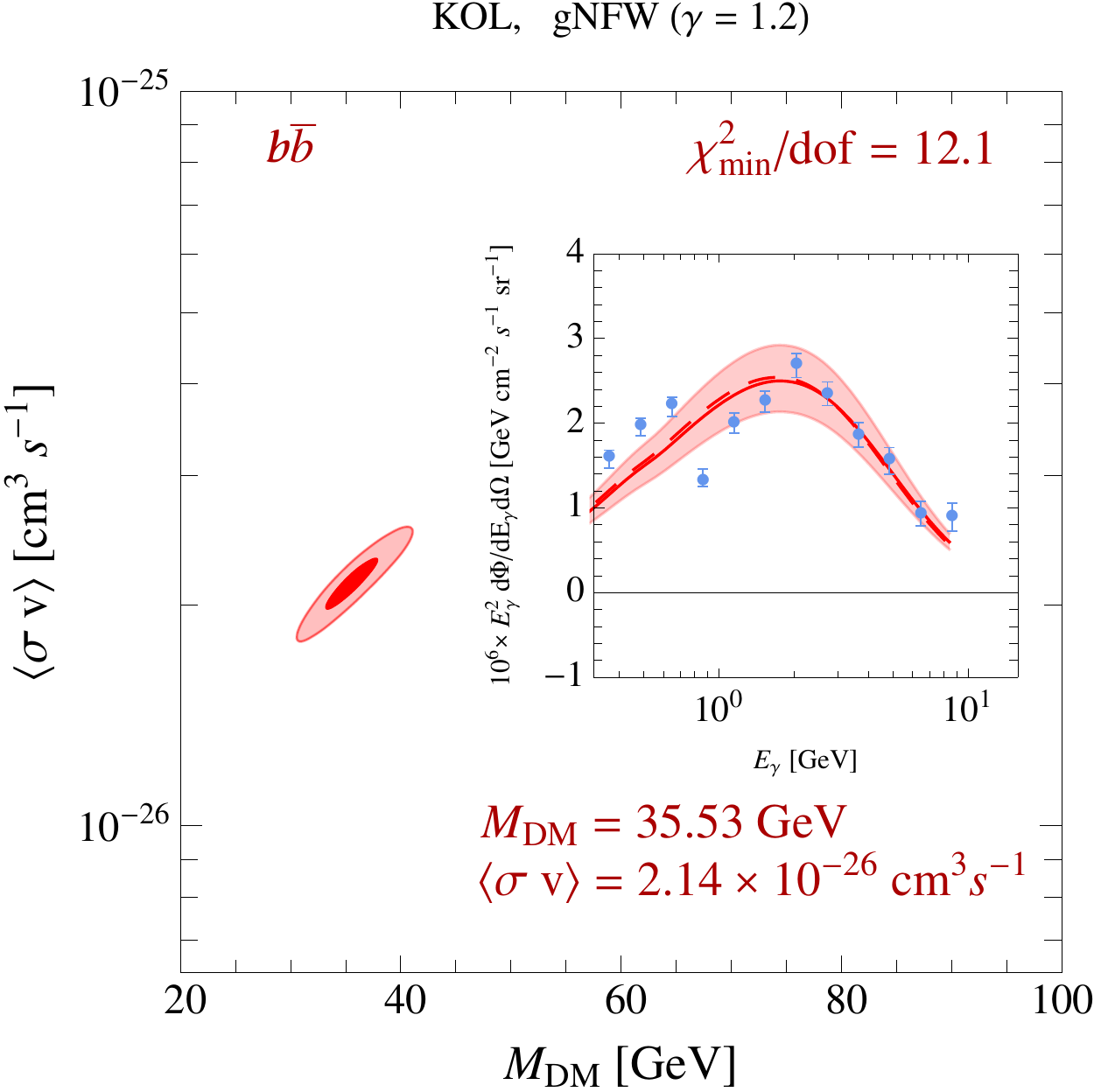}
   %\caption{\textit{Count Map}}\label{fig:CountMap}
    \end{minipage}\hspace{0.3 cm}
   \begin{minipage}{0.45\textwidth}
    \centering
    \includegraphics[scale=0.61]{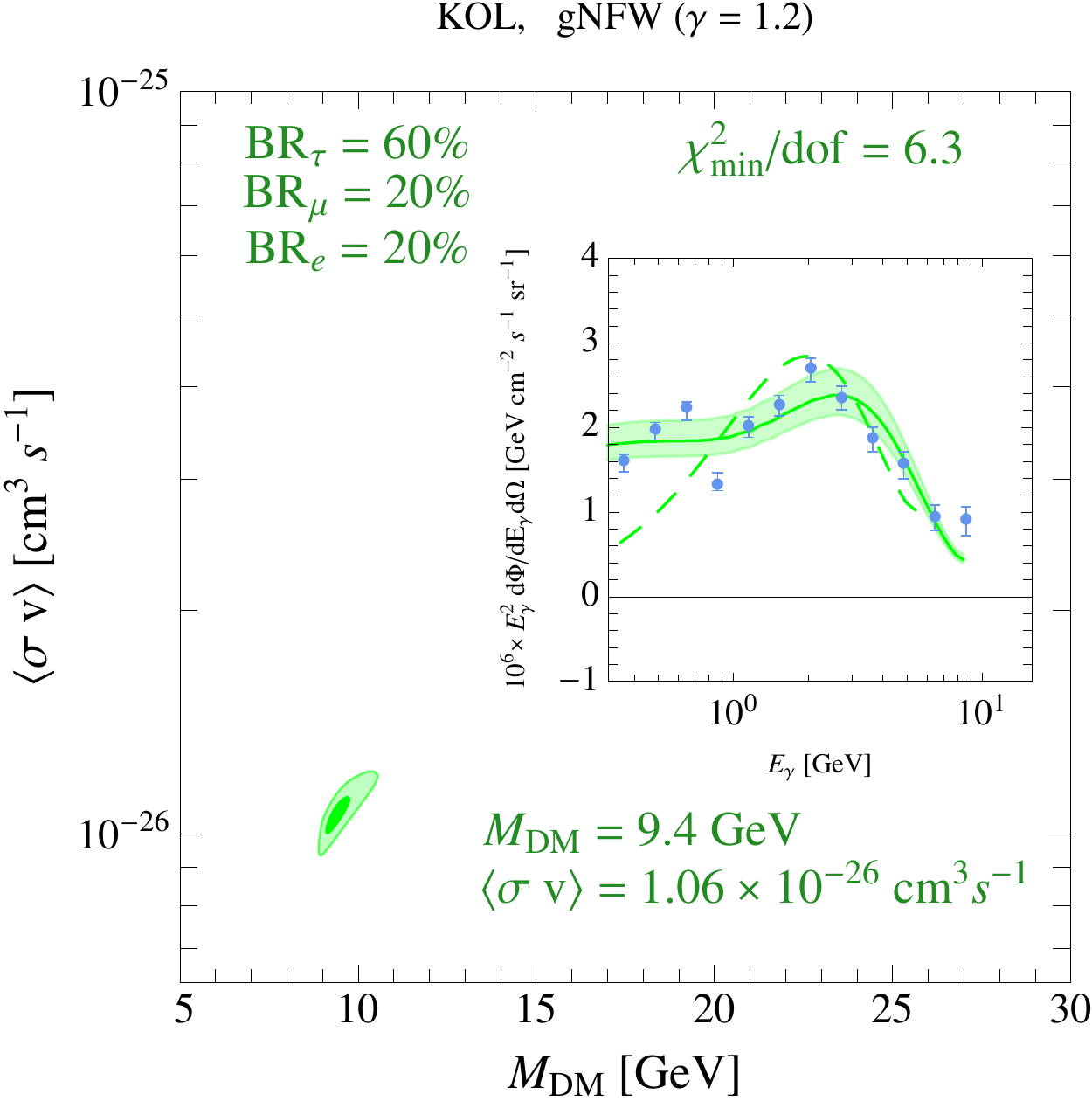}
    \end{minipage}\\[0.5cm]
  \begin{minipage}{0.45\textwidth}
   \centering
   \includegraphics[scale=0.61]{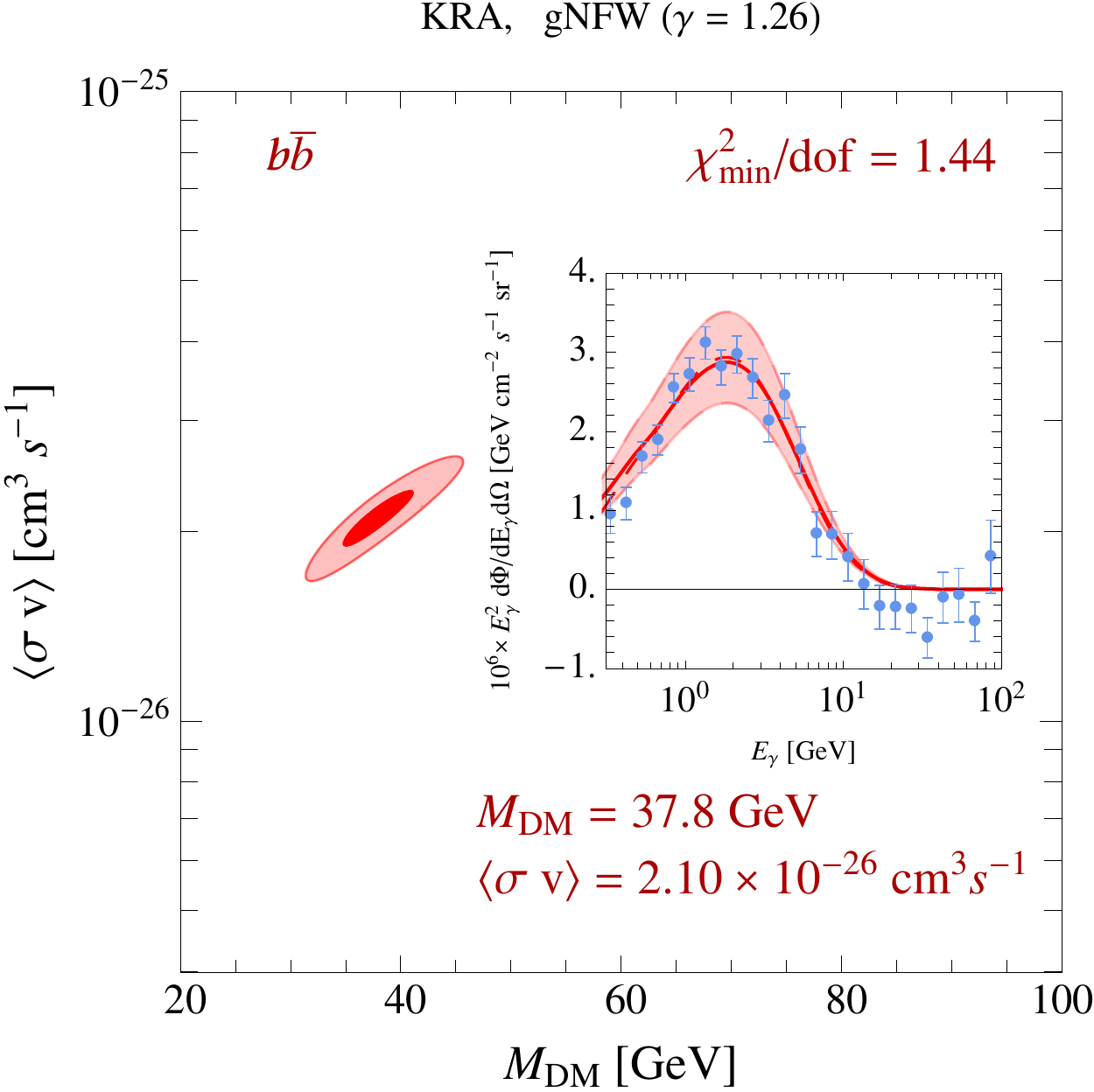}
   %\caption{\textit{Count Map}}\label{fig:CountMap}
    \end{minipage}\hspace{0.3 cm}
   \begin{minipage}{0.45\textwidth}
    \centering
    \includegraphics[scale=0.61]{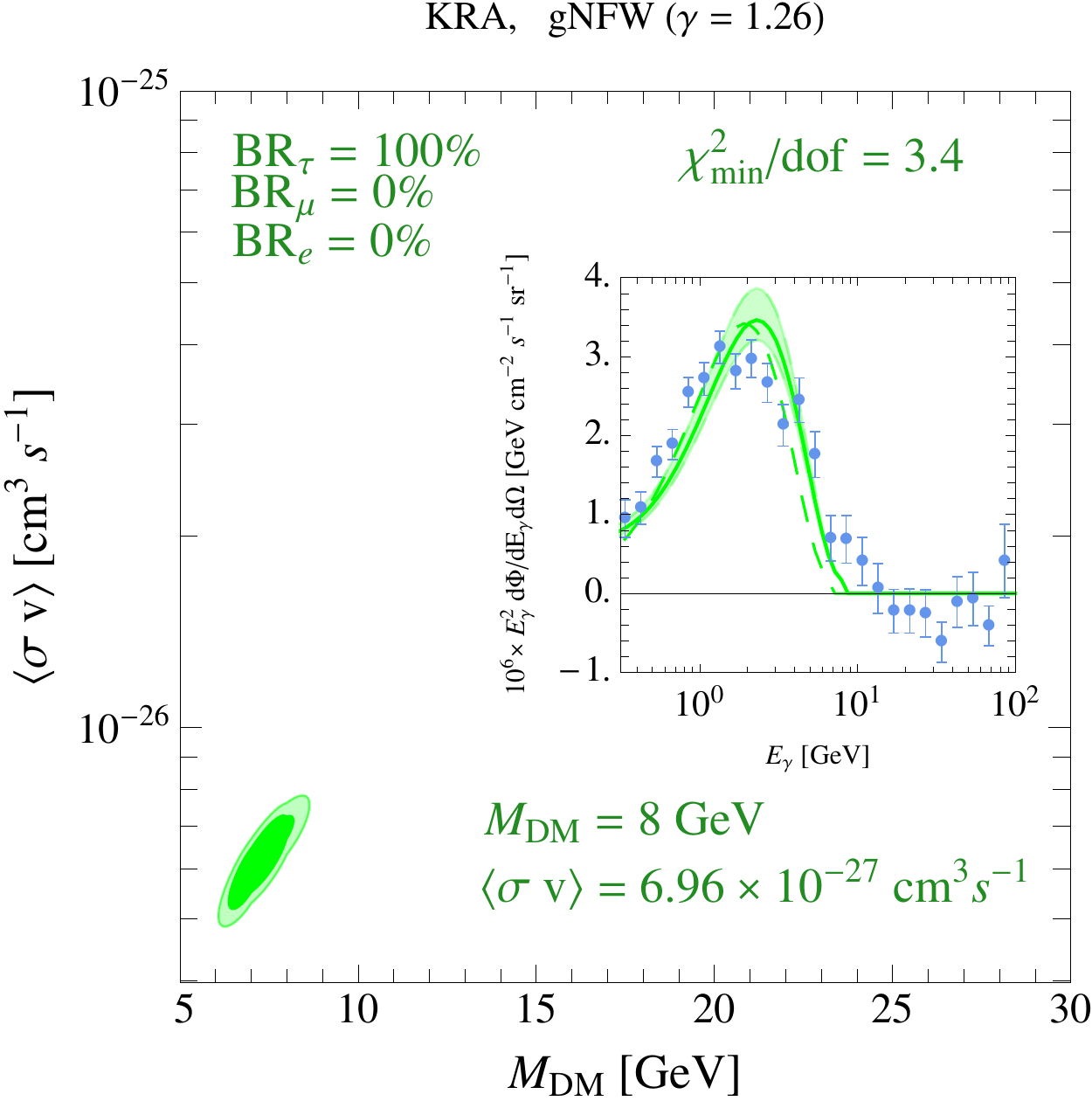}
    \end{minipage}\\
    \vspace{0.25cm}
\caption{\textit{
$\chi$-square fit of the GC excess. We include secondary emissions from DM annihilation, and we show the 1-$\sigma$ and 3-$\sigma$ confidence regions corresponding to 100\% DM annihilation into $b\bar{b}$ (left panels) and DM annihilation into leptons (right panels). 
In the inset plot, we compare the best-fit gamma-ray spectrum with the residual data; for illustrative purposes, the shaded region represents the 3-$\sigma$ band obtained by varying the annihilation cross-section in the corresponding confidence interval (but keeping $M_{\rm DM}$ fixed to the best-fit value).
As far as the $b\bar{b}$ final state is concerned, secondary emissions do not play a significant role.
Considering DM annihilation into leptons, on the contrary, the inclusion of secondary emissions 
can significantly improve the goodness of the fit; 
for comparison, in the inset plots the dashed lines represent the best-fit spectra obtained considering only the prompt emission.
}}
\label{fig:ExSpectrum}
\end{center}
\end{figure}

We show the impact of secondary emissions in fig.~\ref{fig:BBSpectra} for $b\bar{b}$ final state (upper panels), and annihilation into one of the leptonic channels (lower panels). We consider, in each case, two different values for the DM mass, and we explore two different propagation setups, namely the KOL (thicker lines) and the THN (thinner lines) models.
One sees that, as expected, secondary emissions are relevant for the leptonic channel already at an energy corresponding to a fraction of the DM mass. For the $b \bar b$ channel, instead, secondary emissions affect more marginally the spectrum and at smaller energies.

\medskip

In fig.~\ref{fig:ExSpectrum}, we show the results of our fits to GC excess including secondary emissions from DM annihilation, and table~\ref{tab:result} reports the results of our $\chi$-square analysis. We fit the `Galactic Center' data found in ref.~\cite{Daylan:2014rsa} (fig.~7 of ref.~\cite{Daylan:2014rsa}, left panel) and the `Inner Galaxy' data (fig.~5 in ref.~\cite{Daylan:2014rsa}, right panel).  
Concerning the `Galactic Center' analysis, we see that our exemplar leptonic channel with 60\% $\tau^+\tau^-$, 20\% $\mu^+\mu^-$ and 20\% $e^+e^-$ provides a better fit to the data (upper right panel of fig.~\ref{fig:ExSpectrum}), mainly thanks to the low energy tail provided by secondary emissions. This is consistent with the findings of~\cite{Lacroix:2014eea} and their results are therefore confirmed by our more accurate (in terms of energy losses and computation of the emissions) analysis. On the other hand, the shape of the spectrum of the `Inner Galaxy' analysis selects the $b \bar b$ channel as a better fitting possibility, although a 100\% DM annihilation into $\tau^+\tau^-$ can also provide a decent fit (lower right panel of fig.~\ref{fig:ExSpectrum}). 
%Playing with the BRs among the leptons would not sensibly ameliorate the fit. 
One could also consider mixed hadronic/leptonic channels (dubbed e.g. `democratic fermions'), but the qualitative conclusions would remain the same. Changing the propagation setup would change these results only marginally. For completeness, we anyway specify in Table~\ref{tab:result} the employed setup.

\medskip

In summary, this part of our analysis does not want to exhaust or systematically scan all the fitting possibilities. However it illustrates an important point: the specific conclusions that one can draw on the nature of the DM particle responsible for the excess have a critical dependence on (i) the method of extraction of the data and (ii) an accurate computation of the DM gamma-ray flux (including secondary radiation). On the other hand, the choice of the propagation setup for electrons and positrons at the origin of the secondary emissions has a small impact.
%In the future, an important effort should be made to bla bla bla.

\begin{table}[!t]
\centering
\begin{tabular}{|c|c|c|c|c|c|}\hline
     \textbf{Analysis} & \textbf{Final State}  & \textbf{Setup} & $M_{\rm DM}$ [GeV] & $\langle \sigma v\rangle$ [cm$^{3}$\,s$^{-1}$] & $\chi^2_{\rm min}/{\rm dof}$ \\ \hline
          \multirow{2}{*}{`Gal Center', $\gamma = 1.2$} & $b\bar{b}$   & KOL   & 35.53 & $2.14 \times 10^{-26}$ & 12.1 \\ \cline{2-6}
            & leptonic mix\,$^{(\star)}$ & KOL   & 9.4 & $1.06\times 10^{-26}$ & 6.3 \\ \hline
            \multirow{2}{*}{`Inner Gal', $\gamma = 1.26$} & $b\bar{b}$   & KRA   & 37.8 & $2.10 \times 10^{-26}$ & 1.44 \\ \cline{2-6}
           &  $\tau^+\tau^-$ & KRA   & 8 & $6.96\times 10^{-27}$ & 3.4 \\ \hline

                        \end{tabular}\\[0.1cm]
                        \begin{flushright} $^{(\star)}$ leptonic mix = $20\%\ e^+e^-$ + $20\%\ \mu^+\mu^-$ + $60\%\ \tau^+\tau^-$ \phantom{xxx} \end{flushright}
\caption{\label{tab:result} \textit{Results of the $\chi$-square analysis for the fit of the GC excess. }}
     \end{table}

%%%%%%%%%%%%%%%%%%%%%%%%%%%%%%%%%%%%%%%%%%%%%%%%%%%%%%%%%
%%%%%%%%%%%%%%%%%%%%%%%%%%%%%%%%%%%%%%%%%%%%%%%%%%%%%%%%%

\section{Antiproton bound on $b\bar{b}$ final state}
\label{sec:antiproton bounds}

In light of the results of the previous section, it becomes crucial to investigate the DM interpretation of the GC excess from a different but complementary perspective. DM annihilation into $b\bar{b}$ final state copiously produces antiprotons giving rise, in principle, to a detectable signal on Earth. Leptonic annihilation channels, on the contrary, do not feature this property (at least considering values of the DM mass smaller than $\sim 100$ GeV where electroweak radiative corrections -- otherwise able to produce antiprotons via emission of $W^{\pm}$, $Z$ bosons -- play no significant role \cite{Kachelriess:2009zy}). This means that the measurement of the antiproton flux provides a powerful way to scrutinize the hadronic interpretation of the GC excess. To achieve this goal, however, one has to rely on a careful understanding of the astrophysics involved.

%The antiprotons produced by CRs colliding with the interstellar gas, whose production and propagation can be computed with numerical or semi-analytical codes, fit very well -- as already discussed in section~\ref{sec:CR} -- the flux reaching Earth and measured by balloon and satellite experiments like PAMELA. 

\medskip

\begin{figure}[!t]
\begin{center}
\centering
  \begin{minipage}{0.45\textwidth}
   \centering
   \includegraphics[scale=0.6]{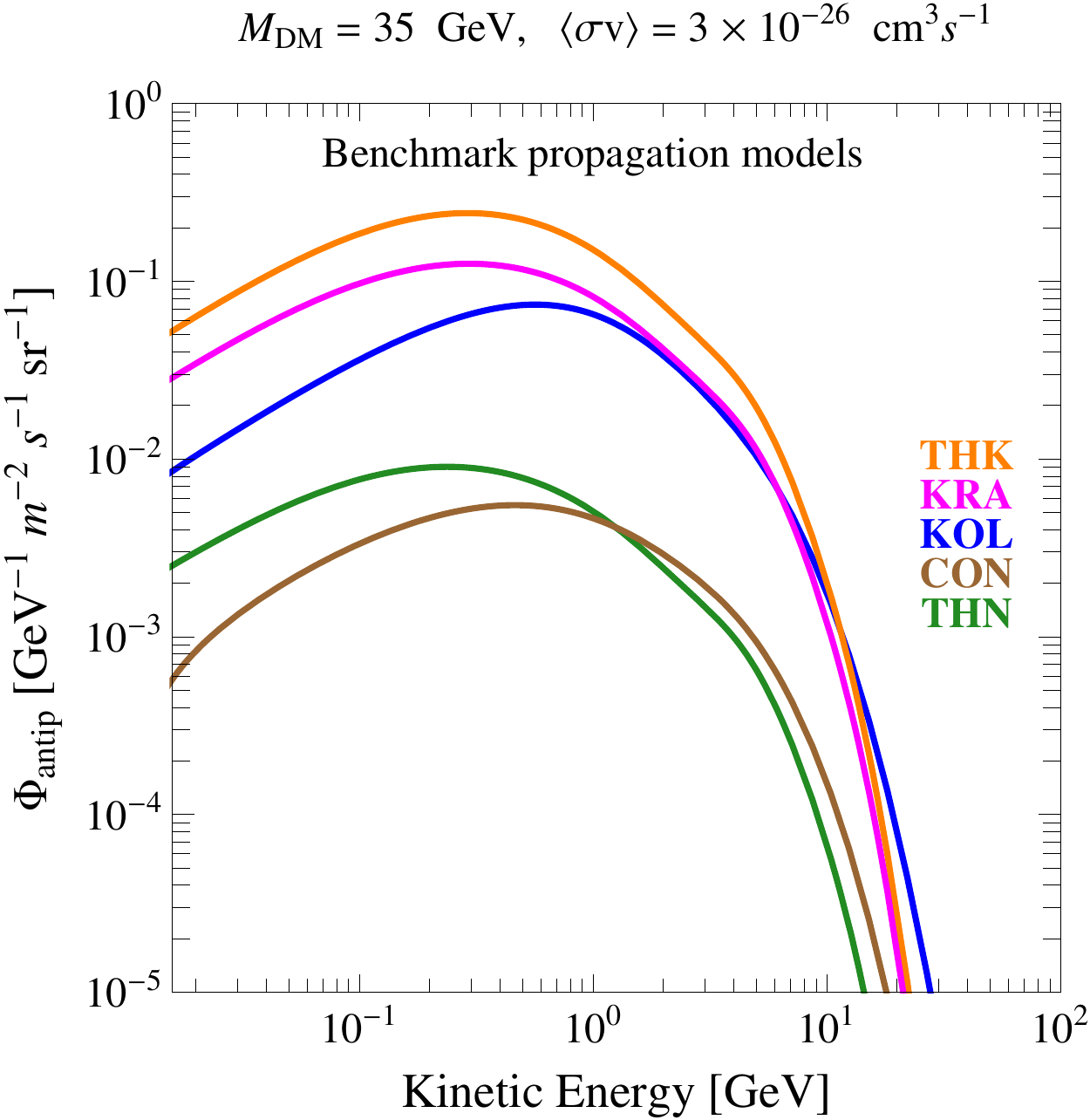}
   %\caption{\textit{Count Map}}\label{fig:CountMap}
    \end{minipage}\hspace{0.1 cm}
   \begin{minipage}{0.45\textwidth}
    \centering
    \includegraphics[scale=0.6]{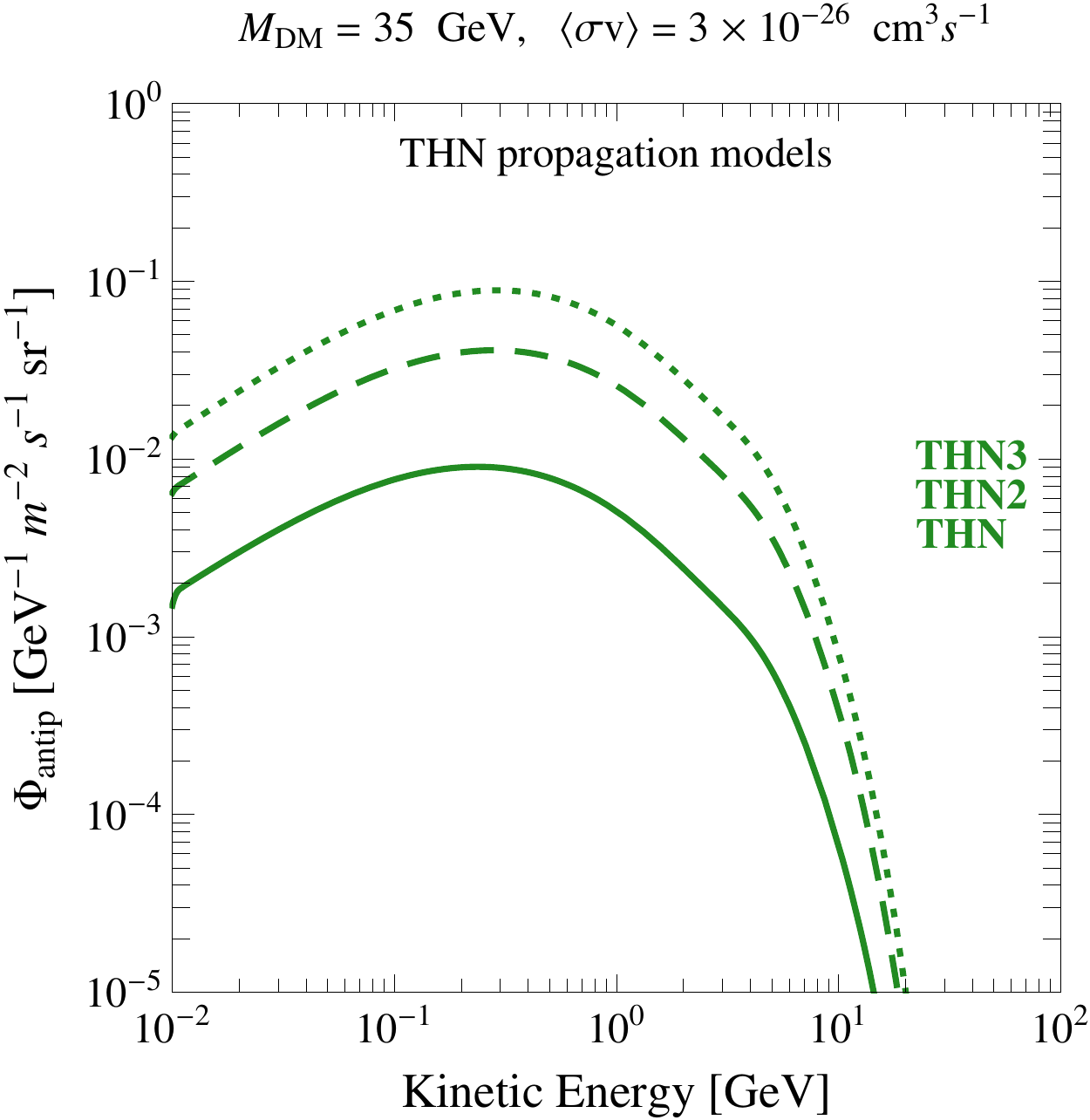}
    \end{minipage}\\
\caption{\textit{
Antiproton flux from DM annihilation into $b\bar{b}$ final state with $M_{\rm DM}= 35$ GeV and $\langle\sigma v\rangle = 3\times 10^{-26}$ cm$^{3}$\,s$^{-1}$. We show the flux at the location of the Earth, after propagation in the Galaxy. We use the gNFW profile with $\gamma = 1.26$.
In the left panel we use the five benchmark propagation models defined in ref.~\cite{Evoli:2011id}. In the right panel we explore 
alternative choices for the scale height $z_t$ that defines the THN model (THN2, dashed; THN3, dotted, see table~\ref{tab:Phi} and text for details).
}}
\label{fig:DMFlux}
\end{center}
\end{figure}

Let us summarize the main points of our approach. First, following the discussion outlined in section~\ref{sec:CR}, we compute the astrophysical antiproton flux originated from $p$-$p$ and $p$-He collisions, for each one of the propagation setups defined in table~\ref{tab:Phi}. Second, we  compute the antiproton flux from DM annihilations assuming 100\% annihilation into $b\bar{b}$ final state, and we propagate it according to the same setup (we plot example fluxes in fig.~\ref{fig:DMFlux}). We assume a gNFW density profile with $\gamma=1.26$ (see eq.~(\ref{eq:gNFW})).   We then constrain this additional DM contribution using the recent {\sc Pamela} data. Since we start from a best fit of the antiproton flux, the constraints are not the most conservative that one can get (more conservative upper limits on DM models may come making no assumption at all about the background astrophysical fluxes) but turn out to be very realistic. 
In practice, we compute the total antiproton flux
\begin{equation}\label{eq:totflux}
\Phi_{\bar p}(M_{\rm DM},\langle \sigma v\rangle,\phi_F^{\bar{p}}) = \Phi_{\bar p}^{\rm BG}(\phi_F^{\bar{p}}) +  \Phi_{\bar p}^{\rm DM}(M_{\rm DM},\langle \sigma v\rangle,\phi_F^{\bar{p}})~, 
\end{equation}
and, using the antiproton data in ref.~\cite{Adriani:2012paa}, we extract a 3-$\sigma$ exclusion contour in the plane $(M_{\rm DM},\langle \sigma v\rangle)$.  

\begin{figure}[p]
\begin{center}
\vspace{-1.0cm}
\fbox{\footnotesize $\phi_{F}^{\bar{p}} = \phi_F^{p}$ fixed}
\end{center}
\begin{center}
\vspace{-0.48cm}
\centering
  \begin{minipage}{0.45\textwidth}
   \centering
   \includegraphics[scale=0.54]{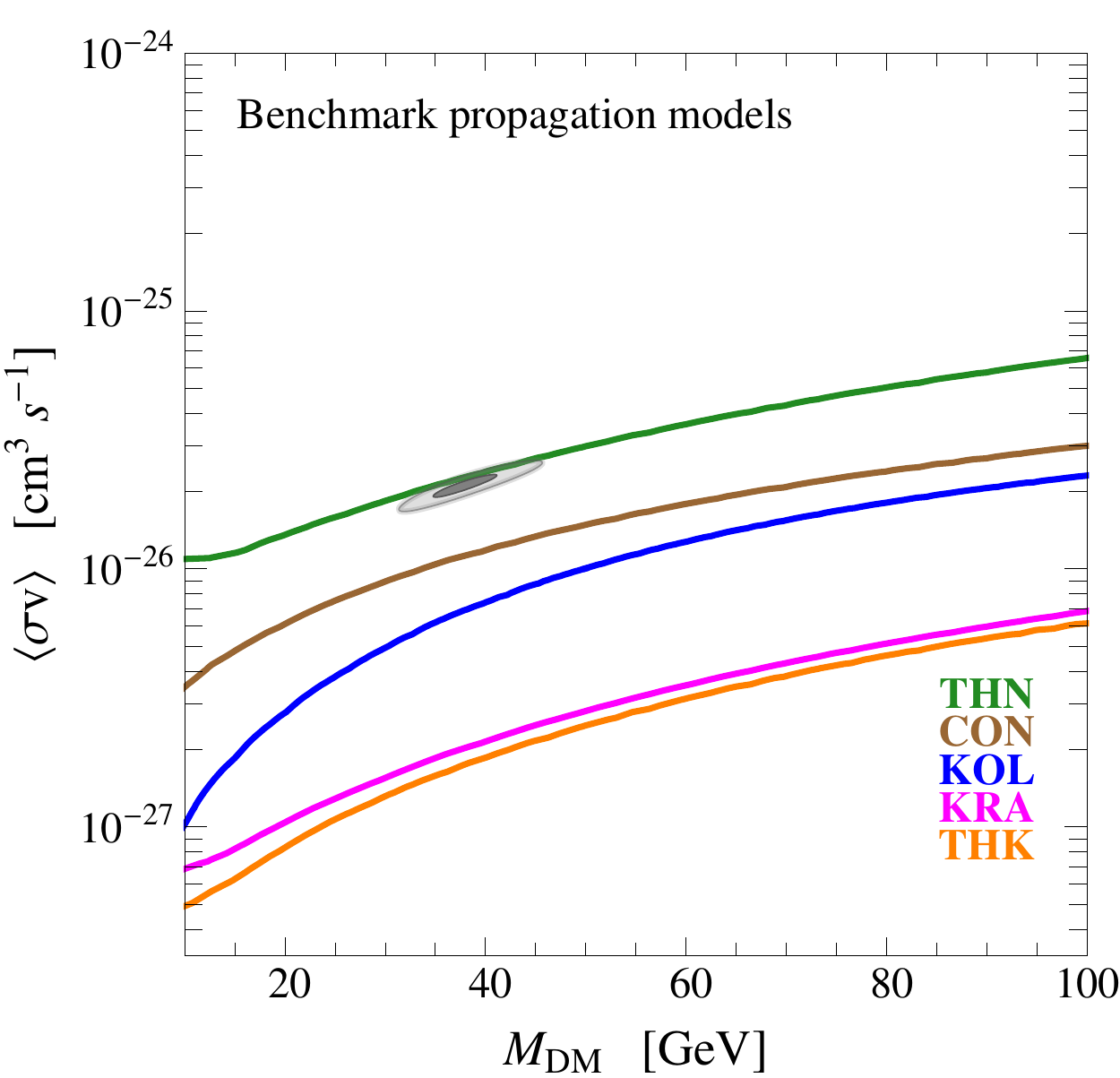}
    \end{minipage}\hspace{0.1 cm}
   \begin{minipage}{0.45\textwidth}
    \centering
    \includegraphics[scale=0.54]{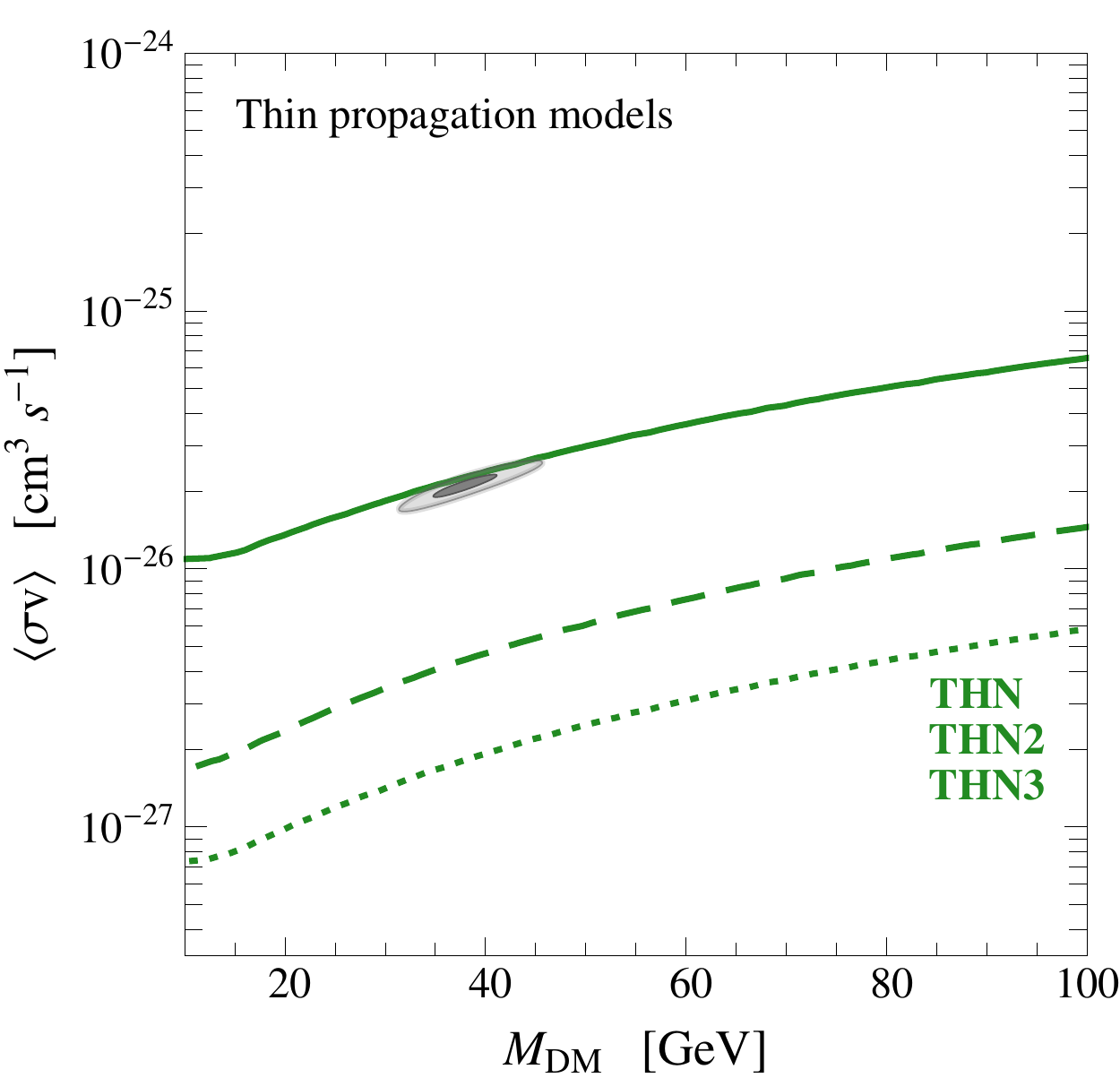}
    \end{minipage}\\
\vspace{0.30cm}
\fbox{\footnotesize $\phi_{F}^{\bar{p}} = \phi_F^{p} \pm 50\%$}
\end{center}
\begin{center}
\vspace{-0.48cm}
\centering
  \begin{minipage}{0.45\textwidth}
   \centering
   \includegraphics[scale=0.54]{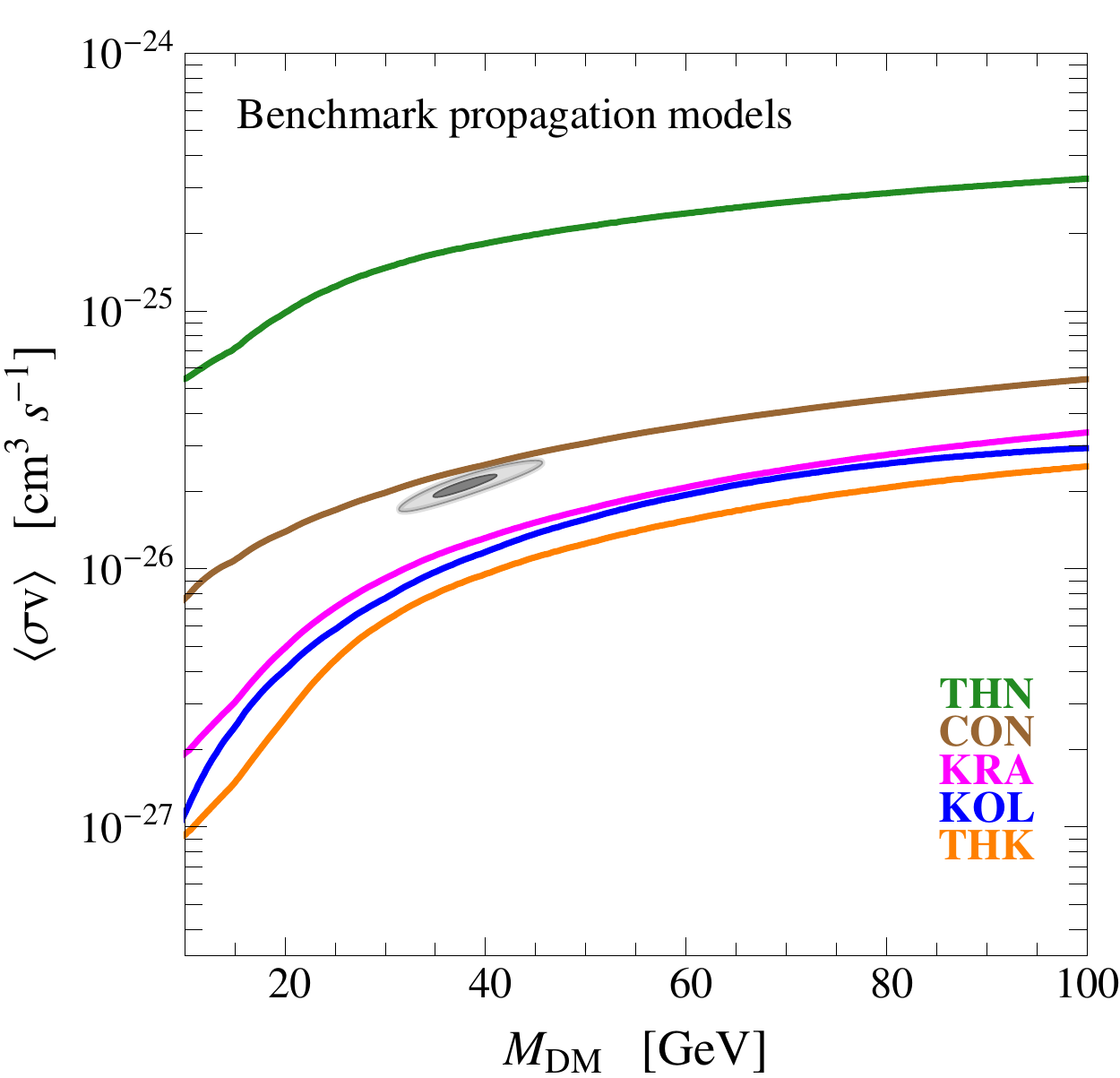}
    \end{minipage}\hspace{0.1 cm}
   \begin{minipage}{0.45\textwidth}
    \centering
    \includegraphics[scale=0.54]{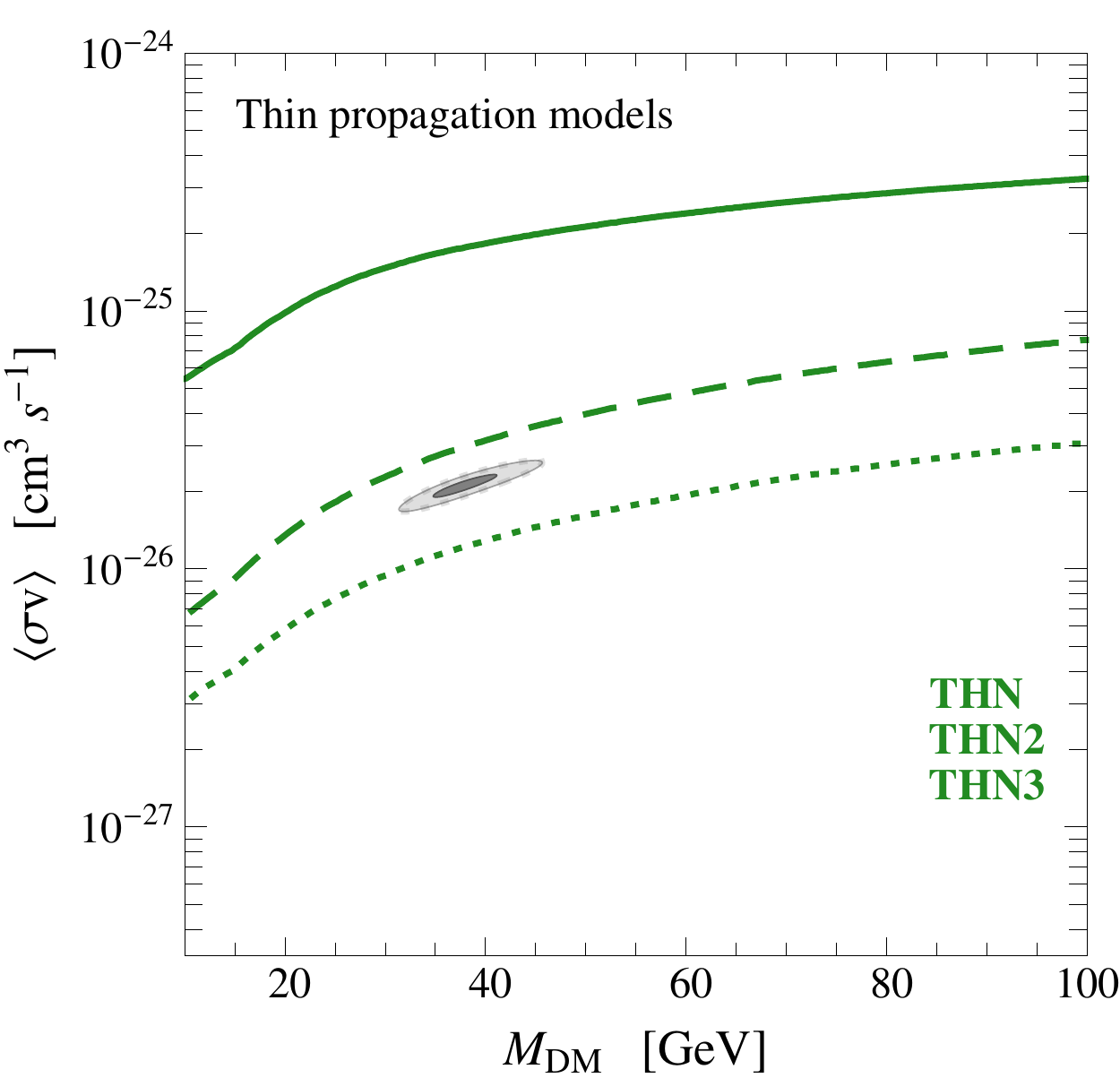}
    \end{minipage}\\
\vspace{0.30cm}
\fbox{\footnotesize $\phi_{F}^{\bar{p}} \in [0.1, 1.1]$ GV}
\end{center}
\begin{center}
\vspace{-0.48cm}\centering
  \begin{minipage}{0.45\textwidth}
   \centering
   \includegraphics[scale=0.54]{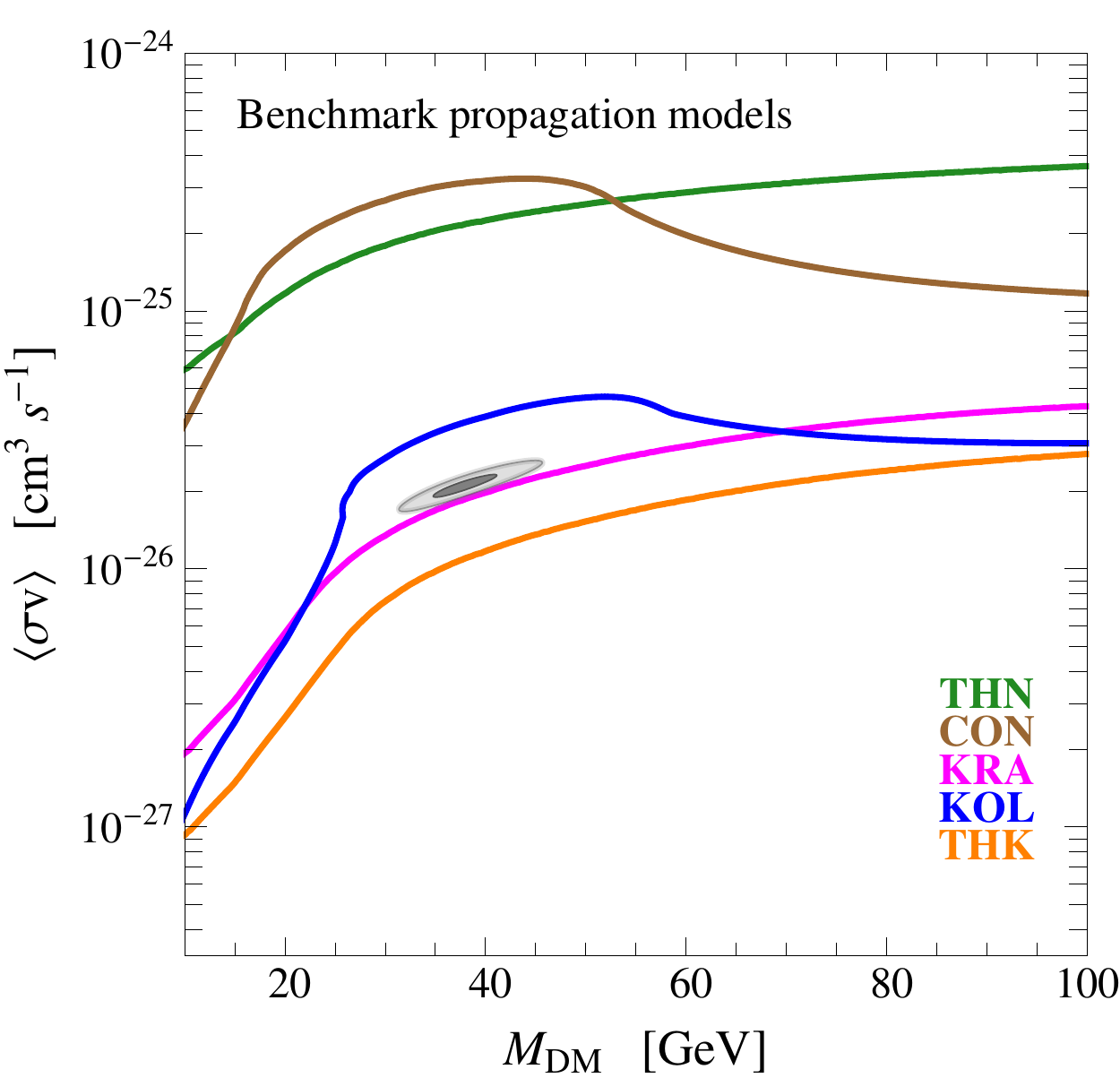}
    \end{minipage}\hspace{0.1 cm}
   \begin{minipage}{0.45\textwidth}
    \centering
    \includegraphics[scale=0.54]{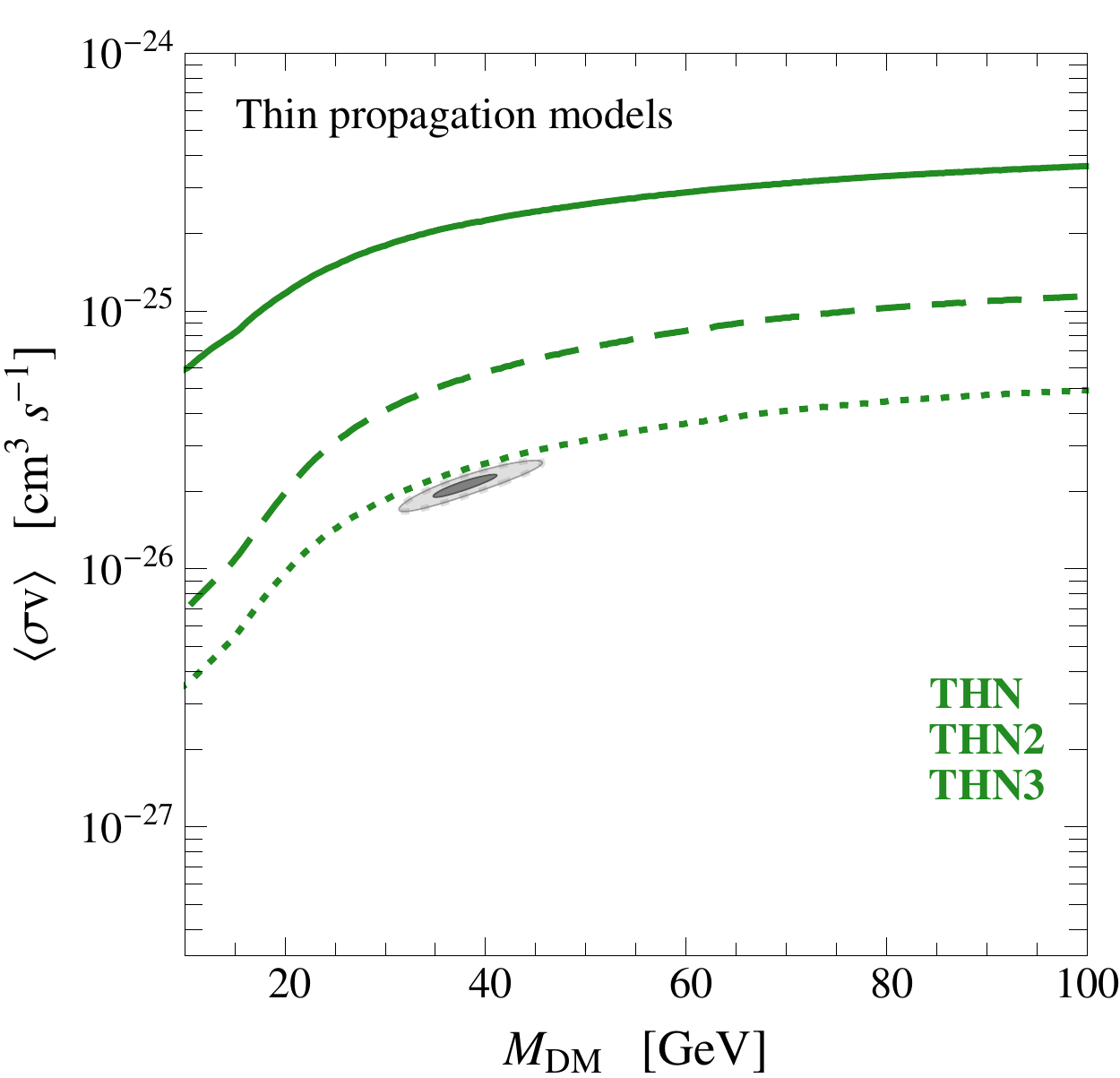}
    \end{minipage}
\caption{\label{fig:XsectionBound}\textit{3-$\sigma$ exclusion contours on $\langle \sigma v \rangle$ for 100\% DM annihilation into $b\bar{b}$, for the three approaches to solar modulation discussed in the text. Left panels: the five benchmark propagation setups. Right panels: alternative choices for the scale height $z_t$ that defines the THN setup (THN2, dashed; THN3, dotted). The gray area is the best-fit region identified in Sec.~\ref{sec:fits}.
}}
\end{center}
\end{figure}    

\medskip

As pointed out in section~\ref{sec:CR}, one crucial ingredient in this kind of analysis is the impact of solar modulation on the total antiproton flux, parametrized in eq.~(\ref{eq:totflux}) by the Fisk potential $\phi_F^{\bar{p}}$ (see eq.~(\ref{eq:Fisk})). In this work, we aim to give an accurate discussion of this delicate issue, and we explore three different approaches:
\begin{itemize}
\item[{\bf A.}] We fix the solar modulation potential $\phi_F^{\bar{p}}$ to the value obtained from the fit of the proton data (see table~\ref{tab:Phi}); we show the corresponding result in the top panels of fig.~\ref{fig:XsectionBound}. 
\item[{\bf B.}] We marginalize over $\phi_F^{\bar{p}}$ considering the interval $\phi_F^{\bar{p}} = \phi_F^{p} \pm 50$\%; we show the corresponding result in the middle panels of fig.~\ref{fig:XsectionBound}.
\item[{\bf C.}] We marginalize over a wider range of values, namely $\phi_F^{\bar{p}} \in [0.1,1.1]$; we show the corresponding result in the lower panels of fig.~\ref{fig:XsectionBound}.
\end{itemize}
Let us now motivate these choices, and discuss in detail our results.

\paragraph{{\bf A.}}Choosing $\phi_F^{\bar{p}} = \phi_F^{p}$ is a very constrictive assumption, and it corresponds to neglecting completely any charge-dependent effect in the description of solar modulation. 
The results obtained in this setup, as a consequence, must be taken with a pinch of salt and lead to the most stringent bound one can obtain from the antiproton data, since there is no freedom to vary the Fisk potential in order to counterbalance -- in particular at low energies -- the impact of  DM on the total flux. In the top left panel of fig.~\ref{fig:XsectionBound} all the propagation setup but THN rule out the DM explanation of the GC excess, represented by the 1- and 3-$\sigma$ confidence regions shaded in gray. 
On the other hand, the THN model, based on a thin diffusion zone with $z_t = 0.5$ kpc, is strongly disfavored by recent studies on synchrotron emission, radio maps and low energy positron spectrum.
%Complementary studies on the synchrotron emission of the Galaxy, in fact, can put constraints on the halo height, whose key role is well pointed out in this plot. 
For example in \cite{DiBernardo:2012zu} the authors claim a 5-$\sigma$ exclusion for models with $z_t < 2$ kpc, with the best fit obtained from both synchrotron profiles and spectra being located around $z_t \simeq 8$ kpc. 
Other works show similar results~\cite{Bringmann:2011py,Orlando:2013ysa,Fornengo:2014mna}: e.g. in \cite{Fornengo:2014mna} the authors model several radio maps of the Galaxy as superpositions of an isotropic component plus the Galactic synchrotron emission, and obtain a preference for large halos in agreement with \cite{DiBernardo:2012zu}.

Therefore, taking into account these results, the exclusion plots shown in the upper left part of fig.~\ref{fig:XsectionBound}  turn out to strongly disfavor the DM interpretation of the gamma-ray excess.

In order to assess more carefully the validity of this claim, we explore in the upper right panel of fig.~\ref{fig:XsectionBound} different choices for the halo height that defines the THN model. 
We compute the antiproton bound corresponding to the THN2 ($z_t = 2$ kpc) and THN3 ($z_t = 3$ kpc) propagation models and we see that the DM interpretation of the GC excess is excluded for both of them.

In summary, within this approach the DM interpretation of the excess requires unrealistically thin ($\lesssim 1$ kpc) diffusive halos, which are at odds with several other measurements, and is therefore strongly disfavored.

\paragraph{{\bf B.}}As mentioned at the beginning, however, this strong conclusion cannot be the ending point of a careful analysis. Protons and antiprotons are likely to be subject to different modulations under the influence of the Sun, and the requirement $\phi_F^{\bar{p}} = \phi_F^{p}$ will be seldom realized. 
For this reason, let us abandon the condition $\phi_F^{\bar{p}} = \phi_F^{p}$, and investigate how the antiproton bound changes as soon as solar modulation is marginalized away in the fit.

In the middle portion of fig.~\ref{fig:XsectionBound}, we marginalize over $\phi_F^{\bar{p}}$ considering the interval $\phi_F^{\bar{p}} = \phi_F^{p} \pm 50$\% (we recall that $\phi_F^{p}$ is fixed for each setup, values are given in table~\ref{tab:Phi}).
We chose this particular range based on the use of the dedicated numerical code {\tt HelioProp}. We considered a range of heliospheric propagation setups in which:
\begin{itemize}
\item[-] The polarity of the SMF and the parameter describing the angular extent of the heliospheric current sheet, namely the {\it tilt angle}, were fixed to the appropriate values for the {\sc Pamela} data taking period. Since the data are dominated by the exceptional 2008-2009 solar minimum, a low value of the tilt angle is expected. In agreement with \cite{Strauss:2012zza} we considered a value of $10^{\circ}$, which was also used in \cite{Gaggero:2013nfa} to model leptonic data.
\item[-] The mean free path in the heliosphere (a free parameter which is generally fit to the data) was allowed to vary in a quite extreme range, from 0.01 AU to 0.4 AU.
\item[-] Also other parameters such as the normalization of the magnetic field and the ratio between perpendicular and parallel diffusion coefficients were varied within very wide ranges. 
\end{itemize}
Then, for every {\tt HelioProp} run we found the Fisk potential that provided the best fit of the modulated  proton and antiproton spectra separately, and found that the relative difference between the value required for $p$ and $\bar{p}$ never exceeded $50$\%, hence the chosen interval for $\phi_F^{\bar{p}}$.

By comparing the results in the middle panels of fig.~\ref{fig:XsectionBound} -- obtained with this range -- with the upper panels, it is evident that in this case the antiproton bound becomes less stringent. Intuitively, this happens since as soon as the DM contribution becomes large enough to overshoot the low-energy data, it can be compensated by increasing the value of the Fisk potential. Nevertheless, we find that also in this case the THK, KOL and KRA propagation models as well as the THN3 model (with $z_t = 3$ kpc) rule out the $b\bar{b}$ DM interpretation of the signal. The bounds corresponding to the CON propagation model and the THN2 model (with $z_t = 2$ kpc), on the contrary, are weakened to the extent that the best-fit region of the GeV excess falls into the allowed region.

\paragraph{{\bf C.}} Finally, in the lower panels of fig.~\ref{fig:XsectionBound} we present even more conservative bounds. Here we choose to marginalize over $\phi_F^{\bar{p}}$ in the interval $\phi_F^{\bar{p}} \in [0.1, 1.1]$ GV (for all propagation setups, irrespectively of the values for $\phi_F^{p}$ they possess).
%This is presumably one of the most conservative choices one can make in the analysis of the antiproton Fisk potential. 
This is a very generous range that most certainly brackets even extreme variations for this parameter.
%In order to motivate this statement, 
One can look for instance at ref.~\cite{SolarModulation}, in which the authors reconstructed the value of the modulation potential over the period from July 1936 through December 2009: according to that analysis, the value of the potential exceeded our upper limit only during a few very pronounced solar maxima around 1990, 1982 and  1959; in the current century the Fisk potential that they derive reached but never exceeded $1.05$ GV. We remind once again that the Fisk potential is model dependent and the assumption on the local interstellar spectrum has a strong impact on its value; nevertheless, since {\sc Pamela} took data during a period of extremely low modulation, the range we adopt appears extremely conservative anyway.

The bounds obtained under these assumptions are of course much weaker: models with large convection or reacceleration such as CON and KOL do not exclude the DM interpretation, as well as models with halo height lower than $3$ kpc. We notice however that models with large halo height, favoured by synchrotron analyses, still rule it out.

%In fig.~\ref{fig:XsectionBoundMargBis}, we marginalize over $\phi_F^{\bar{p}}$ considering the interval %$\phi_F^{\bar{p}} \in [0.1, 1.1]$ GV. 
%This is presumably the most conservative choice one can assume in the analysis of the antiproton Fisk %potential. 
%Let us motivate this statement more carefully. Antiprotons are modulated in the heliosphere
%because of the magnetic activity of the Sun, and this
%modulation varies according to the 11-year solar cycle. In ref.~\cite{SolarModulation} the authors %reconstructed the value of the
%modulation potential for the period from July 1936 through December 2009 [...]

%%%%%%%%%%%%%%%%%%%%%%%%%%%%%%%%%%%%%%%%%%%%%%%%%%%%%%%%%
%%%%%%%%%%%%%%%%%%%%%%%%%%%%%%%%%%%%%%%%%%%%%%%%%%%%%%%%%

\section{Discussion and Conclusions}
\label{sec:conclusions}

It has been known for a long time that the GC is one of the most promising targets for indirect DM search.
Several papers pointed to a gamma-ray excess in {\sc Fermi-LAT} data from that region in the latest years, and, in particular, a recent analysis~\cite{Daylan:2014rsa} confirm its presence with an incredibly high significance. 
%This excess turns out to be best fit by 31-40 GeV DM particles annihilating into $b\bar{b}$ with $\langle \sigma v\rangle = 1.4$-$2\times 10^{-26}$ cm$^3$\,s$^{-1}$.

In this paper we have discussed two key issues related to this presumed DM detection. The first point is the relevance of the secondary gamma radiation emitted by particles originating from DM annihilation. We use the numerical packages {\tt DRAGON} and {\tt GammaSky} in order to compute in a realistic way this contribution and, once this emission is taken into account, we find the best fit regions corresponding to the gamma excess in the (DM mass - cross section) parameter space. We conclude that the secondary emission is relevant for the leptonic channels in a wide energy range. Hence any conclusion on the DM nature of the signal critically depends on this contribution.

\medskip

The second issue we have analyzed is the possibility of confirming or disproving the DM nature of this excess using the antiproton channel,  which has been shown several times to be powerful for these purposes.
We find that the uncertainties on the propagation model, and in particular on the halo height, play a major role, as expected. Moreover, we have discussed in detail the role of the solar modulation, taking into account possible charge dependent effects whose importance is estimated exploiting the numerical package {\tt HelioProp}.

\medskip

Very recently, the authors of ref.~\cite{Bringmann:2014lpa} have also discussed the antiproton bounds. They find that the antiproton data can be marginally consistent with the GeV excess only if the very conservative MIN model from \cite{Donato:2001ms} is used (a model roughly corresponding to our THN). 
We differ from ref.~\cite{Bringmann:2014lpa} since: {\bf 1)} we consider a comprehensive set of propagation models, including several `thin' models with different halo height, and models with high reacceleration or convection together with others where these effects are less important; {\bf 2)} we fully include the subtleties associated to solar modulation: this turns out to be crucial since the more the Fisk potential for the antiprotons is allowed to vary the less stringent the bounds become. 

\medskip

Our overall conclusions are the following: adopting the most realistic propagation models and well motivated choices for the solar modulation potential, the hadronic ($b \bar b$) DM interpretation for the GeV excess is definitely in strong tension with the antiproton data. Nevertheless, given that our knowledge of CR diffusion both in the Galaxy and in the heliosphere is far from being accurate and complete, there are still conservative choices of the parameters involved that do not result in ruling it out, namely thin halo models and large solar modulation potentials. In addition, leptonic channels can provide good fits to the data (although critically dependent on a proper computation of secondary emissions) and avoid antiproton constraints.

In any case, more precise data, and a deeper understanding of CR propagation and modulation, are required to test convincingly the DM origin of this signal with antiprotons. In the future, data from {\sc Ams-02} will possibly have the power to help in these respects.

\bigskip

\acknowledgments

We thank Carmelo Evoli, Dario Grasso, Luca Maccione, Piero Ullio and Andrea Vittino for useful discussions. MC, GG and MT acknowledge the hospitality of the Institut d'Astrophysi-que de Paris, where part of this work was done.
Funding and research infrastructure acknowledgements: 
\begin{itemize}
\item[$\ast$] European Research Council ({\sc Erc}) under the EU Seventh Framework Programme (FP7 2007-2013)/{\sc Erc} Starting Grant (agreement n.\ 278234 --- `{\sc NewDark}' project) [work of MC, GG and MT],
%\item[$\ast$] EU ITN network {\sc Unilhc} [work of M.C.], 
\item[$\ast$] French national research agency {\sc Anr} under contract {\sc Anr} 2010 {\sc Blanc} 041301.
\item[$\ast$] {\sc Erc} Advanced Grant n$^{\circ}$ 267985, `Electroweak Symmetry Breaking, Flavour and Dark Matter: One Solution for Three Mysteries' ({\sc DaMeSyFla}) [work of AU].
 \end{itemize}

\end{document}